\documentclass[nofootinbib,floatfix,twocolumn]{revtex4}

\usepackage{empheq} 
\usepackage{mathrsfs}
\usepackage{verbatim}
\usepackage{amssymb}
\usepackage{graphicx}
\usepackage{subfigure}
\usepackage{color}
\usepackage{rotating}

\usepackage[mathscr]{euscript}
\usepackage{amsmath}

\def\bhx{{\bf {\hat X}}}
\def\bhy{{\bf {\hat Y}}}
\def\bhPhi{{\bf {\hat \Phi}}}
\def\bhPsi{{\bf {\hat \Psi}}}
\def\bhi{{\bf {\hat i}}}
\def\bhjy{{\bf {\hat j}}}
\def\bhk{{\bf {\hat k}}}

\def\br{{\bf r}}

\def\bp{{\bf p}}
\def\bn{{\bf n}}
\def\bl{{\bf L}}
\def\bhl{{\bf {\hat L}}}
\def\bs{{\bf S_{eff}}}
\def\bsone{{\bf S_1}}
\def\bstwo{{\bf S_2}}
\def\bhs{{\bf {\hat S}}}
\def\bj{{\bf J}}
\def\bhj{{\bf {\hat J}}}
\def\bn{{\bf {\hat n}}}

\def\be{\begin{equation}}
\def\ee{\end{equation}}
\def\ba{\begin{eqnarray}}
\def\ea{\end{eqnarray}}

\newcommand{\half}[0]{\frac{1}{2}}

\begin{document}
\bibliographystyle{apsrev}

\title{Dynamics of Black Hole Pairs II:\par Spherical Orbits and the
  Homoclinic Limit of Zoom-Whirliness}

\author{Rebecca Grossman${}^{**}$ and Janna Levin${}^{*,!}$}
\affiliation{${}^{**}$Physics Department, Columbia University,
New York, NY 10027}
\affiliation{${}^{*}$Department of Physics and Astronomy, Barnard
College of Columbia University, 3009 Broadway, New York, NY 10027 }
\affiliation{${}^{!}$Institute for Strings, Cosmology and Astroparticle
  Physics, Columbia University, New York, NY 10027}
\affiliation{ becky@phys.columbia.edu}
\affiliation{janna@astro.columbia.edu}

\widetext

\begin{abstract}

Spinning black hole pairs exhibit a range of complicated dynamical
behaviors. An interest in eccentric and zoom-whirl orbits has
ironically inspired the focus of this paper: the constant radius
orbits. When black hole spins are misaligned, the constant radius orbits are not
circles but rather lie on the surface of a sphere and have acquired
the name ``spherical orbits''. The spherical orbits are significant as
they energetically frame the distribution of all orbits. In addition,
each unstable spherical orbit is asymptotically approached by an orbit
that whirls an infinite number of times, known as a homoclinic
orbit. A homoclinic trajectory is an infinite whirl limit of the zoom-whirl spectrum
and
has a further significance as the separatrix between
inspiral and plunge for eccentric orbits. We work in the context of
two spinning black holes of comparable mass as described in the 3PN
Hamiltonian with spin-orbit coupling included. As such, the results
could provide a testing ground of the accuracy of the PN
expansion. Further, the spherical orbits could provide useful initial
data for numerical relativity. Finally, we comment that
the spinning black hole pairs should give way to chaos
around the homoclinic orbit when spin-spin coupling is incorporated.

\end{abstract}

\maketitle

A complete knowledge of the dynamics of
black hole pairs is essential for future gravitational wave
experiments. Yet the importance of dynamics has not always been
appreciated. Although stellar mass black hole pairs are significant
candidates for a first direct detection with LIGO, 
their detectable gravitational radiation would be 
emitted from nearly circular orbits -- at least that was the refrain.
This
preferential focus on quasi-circular inspiral was motivated by
considerations of long-lived binaries that begin with a modest
eccentricity that is gradually shed as angular momentum is lost to
gravitational waves. A fair assessment of known astrophysics, the
claim discouraged analyses of orbital dynamics in favor of the simpler
analysis of circular orbits.

However, black hole binaries formed by tidal
capture in dense star clusters, such as globular clusters, do not
conform to this story \cite{portegies-zwart1999}. As one black hole 
scatters with another black hole in a dense region,
a burst of radiation is emitted on close
encounter. Some subset of these encounters will leave the pair
bound in a highly eccentric orbit that merges too quickly to
circularize. Estimates conclude that as many as 30\% of multi-black hole
systems will retain eccentricities $>0.1$ as their waves sweep into the LIGO
bandwidth \cite{wen2003}. 

Most recently, a new source of eccentric
mergers was predicted to have a substantial detection rate \cite{O'Leary:2008xt}. Black hole/black
hole scattering in galactic nuclei would similarly lead to tidal
captures and highly eccentric, short-lived black hole binaries with
90\% entering the LIGO bandwidth with eccentricities $>0.9$
\cite{O'Leary:2008xt}. These competitive
sources for a first detection by Advanced LIGO
\cite{O'Leary:2008xt} 
further motivate our study of the complete
dynamics of binary black holes 
\cite{levino2000,levin2000,levin2003}.

In paper I in this series \cite{levin2008:2}, we presented the
spectrum of orbits in the strong-field regime
when only one body spins.\footnote{In this paper we will use the word
  ``orbit'' to mean bound, non-plunging trajectories.}
There we found zoom-whirl
behavior -- during which an orbit zooms out in large leaves followed
by nearly circular inner whirls. Significantly, 
this extreme form of
perihelion precession is prevalent in comparable mass systems
\cite{levin2008:2}, just as it is in extreme-mass-ratio inspirals
\cite{levin2008}.
Zoom-whirl behavior is not restricted to extreme eccentricities, but
can be executed by orbits of all eccentricities in the strong-field.
We should therefore be prepared to detect evidence of such black hole dynamics
in gravitational waves.

\begin{figure}
\hspace{0pt}
  \centering
\includegraphics[width=60mm]{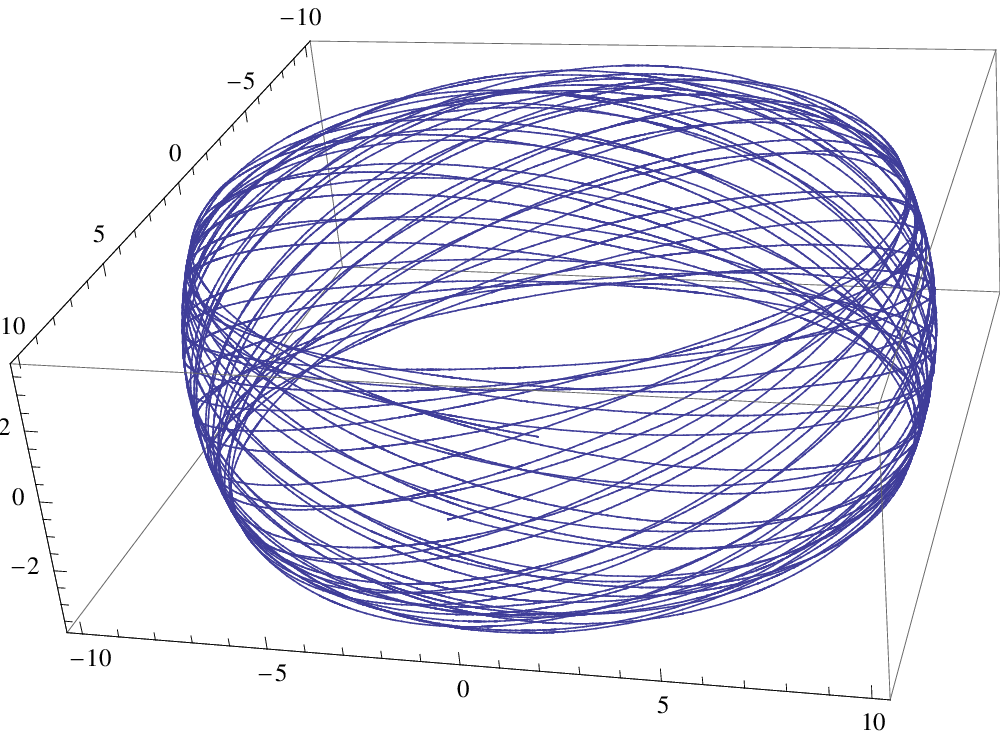}
\includegraphics[width=45mm]{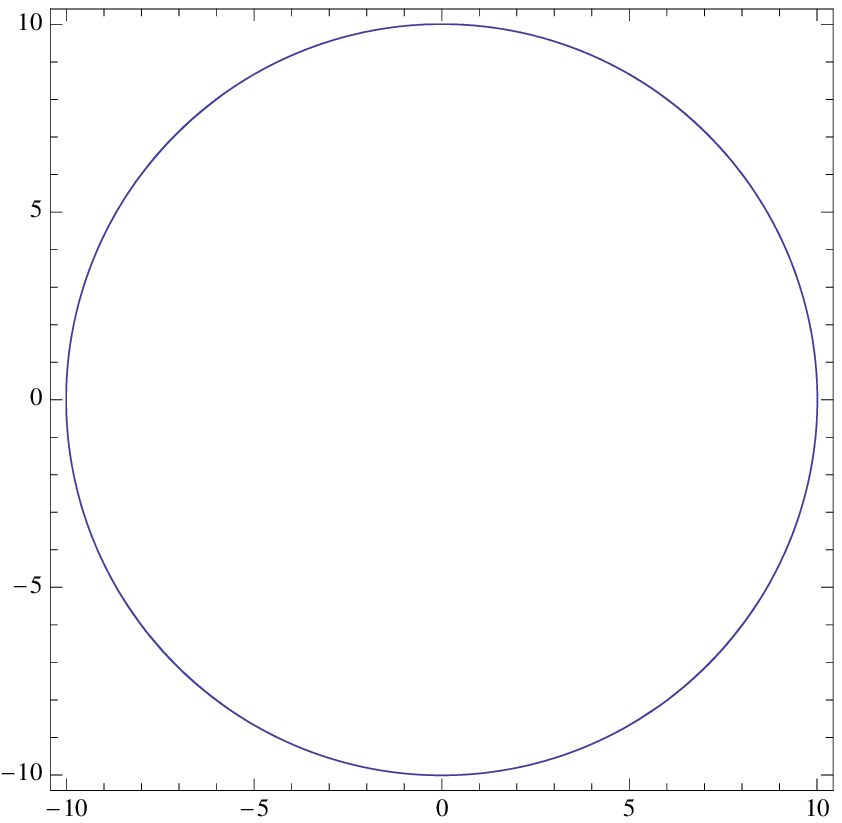}
\hfill
  \caption{A spherical orbit for
mass ratio $m_2/m_1=1/4$ and spin amplitudes of $3/4$. Both spins are initially displaced
    from the orbital angular momentum by $\pi/4$. Notice the orbit is
    not closed. Upper Panel:
  The three-dimensional orbit fills out a strip on a sphere. If we
  waited long enough, the band would be solidly painted, a reflection
  of the aperiodicity of the orbit. Lower Panel: The
  path as caught by the orbital plane reveals the constant radius.
}
  \label{examplesphere}
\end{figure}

In this companion to paper I, we work again in the conservative
Hamiltonian 3PN approximation plus spin-orbit coupling, but 
move beyond
paper I to consider two spinning black holes in a binary.
We focus on
special sets of orbits, namely the spherical orbits. 
That might seem ironic
since
we have just
argued that 
gravitational wave science will probe the full range of dynamical
possibilities, 
there are several good
reasons to devote some time to constant radius orbits. 
\begin{enumerate}
\item If even one black
hole spins, the constant radius orbits are no longer circles (unless
spins are exactly aligned or anti-aligned with the orbital angular
momentum) \cite{damoureob2001,buonanno2006}. As a
result of spin precession, they
fill a band on the surface of a sphere and have thereby been
coined spherical orbits (see Fig.\ \ref{examplesphere}). Long-lived
binaries that have shed enough angular momentum to lose their
eccentricity but not their spin will exhibit quasi-spherical inspiral
and not quasi-circular inspiral. To detect waves from realistic
binaries,
we will need to understand the orbital parameters and precessions of these
spherical orbits. 
\item Black hole pairs that enter the LIGO bandwidth with their
eccentricity intact will evolve through a sequence of zoom-whirl
orbits rather than nearly spherical ones.
Still,  
the spherical sets are special since they mark the minimum and maximum
energy 
in the strong-field
spectrum of bound orbis for a given angular momentum.\footnote{As
  detailed in the paper, we {\it define} the strong-field by the appearance
  of bound, unstable spherical orbits.} The orbital demographic is
therefore entirely determined by the spherical orbits
\cite{levin2008:2}.
\item The energetically-bound unstable spherical orbits mark the divide
between inspiral and plunge \cite{Sperhake:2007gu}. More specifically, black hole
spacetimes harbor homoclinic orbits -- orbits that approach the
unstable spherical orbits in the infinite future or in the infinite
past
\cite{bombelli1992,levino2000,levin2008:3,perez-giz2008}.
A Homoclinic orbit, often refered to as a separatrix, is
a classic feature of a non-linear dynamical system
and deserve attention since they 
mark the
transition from inspiral to plunge for {\it all} pairs.  As we
mention in the close of this paper, when spin-spin coupling is
incorporated in the PN Hamiltonian, the homoclinic set can become the locus of a transition from
regular to chaotic
behavior.
\end{enumerate}

Among the infinite list of spherical orbits, two are valuable enough
to deserve names:
the
innermost stable spherical orbit (isso) and the innermost bound spherical orbit
(ibso). The acronyms are drawn in analogy with the equatorial isco (innermost
stable circular orbit) and ibco (innermost bound circular orbits).
The isso is the lowest energy spherical orbital and the ibso is the
highest energy, bound spherical orbit.

The isco, well-known as the site of
the transition from
inspiral to plunge for quasi-circular inspiral,
is actually the zero eccentricity
homoclinic orbit \cite{bombelli1992}. {\it All other orbits}, besides the quasi-circular
one, will transition to plunge through another member of the
homoclinic family, hence the importance of the homoclinic set to
gravitational wave science \cite{levin2008:3,perez-giz2008}.
To our
knowledge the homoclinic orbits have not yet been identified in the PN
Hamiltonian expansion before this paper, although an earlier paper
found homoclinic orbits and zoom-whirl behavior in a hybrid PN
expansion \cite{levino2000}.
Excitingly enough,
homoclinic orbits have been observed in fully relativistic, numerical
treatments as well
\cite{pretorius2007}. 

Taken together this special set -- composed of spherical and homoclinic orbits -- demarcates 
dynamical regions and
we spend time on their attributes in this paper. Due to the lack of
confidence in the PN expansion at close separations, we do not
advocate that these results be taken as quantitatively accurate
descriptions of binary black hole dynamics, but rather as
qualitatively descriptive.\footnote{The weakness of the PN approximation
famously plagues other attempts to pinpoint the transition from
inspiral to plunge through the isso \cite{buonanno2006}.} Ineed, we
point out peculiar artifacts of the 3PN system as we go along and the
results of this paper could provide a new terrain on which to test the
PN expansion against, for instance, 
numerical relativity.

Our approach has some overlap with, but is not redundant with, the
Refs.\ \cite{{damoureob2001},{buonanno2006}} and
allows us to find homoclinic orbits and
stability exponents for use in the periodic orbit taxonomy 
of paper I \cite{levin2008:2}.
We also simplify the initial conditions for spherical
orbits
in the absence of radiation
reaction. For quasi-spherical orbits with radiation reaction included
see \cite{buonanno2006}.

The outline of the paper is as follows: In \S \ref{hams} we write out
the equations of motion for two spinning bodies in an orbital basis,
relying on the results of appendix \ref{orbitalapp}. 
In \S \ref{spherical} we determine the orbital parameters of spherical
orbits.
In \S \ref{homoclinic} we find the homoclinic orbits and 
emphasize their connection to 
dynamical instability.
In the conclusion, \S \ref{sum}, we discuss the destruction of the spherical orbits
and the transition to chaos when spin-spin coupling is included.

\section{3PN Hamiltonian + SO Coupling}
\label{hams}

We will work with a condensed and revealing set of equations of
motion in a non-orthogonal orbital coordinate system as derived in \cite{levin2008:2}.
For reference in this companion to that paper, we write out the usual
3PN Hamilton plus spin-orbit coupling for two spinning black holes. 

In a Hamiltonian formulation, the equations of motion are derived from
\begin{equation}
\dot {\br}=\frac{\partial H}{\partial {\bp}} \quad ,\quad
\dot{\bp}=-\frac{\partial{H}}{\partial {\br}}\quad .
\label{hameoms}
\end{equation}
As is standard convention, we work in dimensionless coordinates: 
the dimensionless coordinate vector, ${\bf r}$, is measured in units of
total mass, $M=m_1+m_2$, for a pair with black
hole masses $m_1$ and $m_2$.
The canonical
momentum,
 ${\bf p}$, is measured in units of the reduced mass,
$\mu=m_1m_2/M$. The dimensionless combination
$\eta =\mu/M$ will prove useful. We write vector quantities
in bold. The coordinate $r$ is to be understood as the magnitude 
$r=\sqrt{\br\cdot \br}$. Unit vectors such as $\bn = \br/r$ will additionally
carry a hat.
Finally, we have used the 
dimensionless reduced Hamiltonian $H={\cal H}/\mu$ in Eqs.\ (\ref{hameoms}),
where ${\cal H}$ is the physical Hamiltonian, to 3PN
order plus spin-orbit terms 
\cite{{schaefer1985},{damour1981},{damour1988},{jaranowski1998},{damourpn2001},{damourpn2000:2}}. 
$H$ can be expanded as
\begin{equation}
H= H_N+H_{1PN}+H_{2PN}+H_{3PN}+H_{SO} \quad ,
\label{ham1}
\end{equation}
where
\begin{widetext}
\begin{align}
H_N&=\frac{{\bp}^2}{2}-\frac{1}{r} &  & \\
H_{1PN}&=\frac{1}{8}\left (3\eta-1\right ) \left ({\bp}^2\right )^2
-\frac{1}{2}\left [\left (3+\eta\right ){\bp}^2+\eta({\bn} \cdot 
{\bf  p})^2\right ] \frac{1}{r}+\frac{1}{2r^2} &\nonumber \\
H_{2PN}&= \frac{1}{16} \left (1-5\eta+5\eta^2\right ) \left ({\bp}^2\right )^3
+\frac{1}{8}\left [\left (5-20\eta-3\eta^2\right)\left ({\bp}^2\right )^2 \right.&\nonumber \\
&\left.- 2\eta^2({\bn} \cdot {\bp})^2{\bp}^2
-3\eta^2({\bn} \cdot {\bp})^4\right ] \frac{1}{r} &\nonumber \\
& +\frac{1}{2}\left [\left (5+8\eta\right ){\bp}^2
+3\eta({\bn} \cdot {\bp})^2\right ] \frac{1}{r^2}
-\frac{1}{4}\left( 1+3\eta \right)\frac{1}{r^3} &
\nonumber \\
H_{3PN}&=\frac{1}{128}\left (-5+35\eta-70\eta^2+35\eta^3\right )
\left ({\bp}^2\right )^4
+\frac{1}{16}\left [\left (-7+42\eta-53\eta^2-5\eta^3\right )\left
  ({\bp}^2\right )^3\right. &\nonumber \\
& \left. +(2-3\eta)\eta^2({\bn} \cdot {\bp})^2({\bp}^2)^2
+3(1-\eta)\eta^2({\bn} \cdot {\bp})^4{\bp}^2-5\eta^3({\bn} \cdot
{\bp})^6\right ]\frac{1}{r}&\nonumber \\
&+\left [\frac{1}{16}(-27+136\eta+109\eta^2)({\bp}^2)^2
+\frac{1}{16}(17+30\eta)\eta({\bn} \cdot {\bp})^2{\bp}^2 +
\frac{1}{12}(5+43\eta)\eta({\bn} \cdot {\bp})^4\right ]
\frac{1}{r^2}&
\nonumber \\
&+\left \{ \frac{1}{192}\left [-600+\left
  (3\pi^2- 1340 \right  )
\eta-552\eta^2\right ]{\bp}^2 
-\frac{1}{64}\left(340+3\pi^2+112\eta\right )\eta
({\bn} \cdot {\bp})^2\right \} \frac{1}{r^3}&
\nonumber \\
&+\frac{1}{96}\left [ 12+\left (872-63\pi^2\right )
  \eta\right ]\frac{1}{r^4}\quad ,&
\nonumber \\
H_{SO}&=\frac{\bl\cdot \bs}{r^3}  \quad .
\label{Hterms}
\end{align}
\end{widetext}
For two spinning black holes $\bs$ is\footnote{The definitions for
  $\bs$ can vary in the literature up to an overall constant although
  the reduced $H_{SO}$ must be the same for all prescriptions.}
\begin{equation}
\bs=\delta_1\bsone +\delta_2\bstwo
\end{equation}
where the dimensionless reduced spins are defined as
\begin{equation}
\bsone={\bf a_1} (m_1^2/\mu M)\quad , \quad \quad
\bstwo={\bf a_2} (m_2^2/\mu M)\quad .
\end{equation} 
and
\begin{equation}
\delta_1\equiv\left (2+\frac{3m_2}{2m_1}\right )\eta \quad , \quad\quad
\delta_2\equiv\left (2+\frac{3m_1}{2m_2}\right )\eta \quad .
\end{equation}
The dimensionless spin amplitudes are confined to the range
$0\le a_{1,2}\le 1$.
The reduced orbital angular momentum $\bl=\br\times\bp$ and the
spins precess
according to
\begin{eqnarray}
{\bf \dot {\bf L}} &=& \frac{\bs\times \bf L}{r^3} \nonumber \\
{\bf \dot{{\bf S}}_1}  &=& 
\delta_1\frac{\bl\times \bsone}{r^3}\nonumber \\
{\bf \dot{{\bf S}}_2}  &=& 
\delta_2\frac{\bl\times \bstwo}{r^3} 
\quad .
\label{ds2}
\end{eqnarray}
The spin precessions can be grouped together,
\begin{eqnarray}
{\bf \dot{{\bf S}}_1} +{\bf \dot{{\bf S}}_2} &=& 
\frac{\bl\times \bs}{r^3}
\quad .
\label{dL}
\end{eqnarray}
Notice that the precession of the sum of the spins is equal and opposite to the
precession of the orbital angular momentum. So that
$\bj=\bl+\bsone+\bstwo$
is
conserved. The magnitudes $L, S_1$ and $S_2$,
the inner product $\bs\cdot\bl$, and the energy (the 
Hamiltonian) are also constant for a given orbit.

In general, neither
$\bhj\cdot \bhl$ nor the magnitude $|S_{\rm eff}|$ is
constant. However, there are notable exceptions
\cite{gopakumar2005}.
Both $\bhj\cdot \bhl$ and the magnitude $|S_{\rm eff}|$ are
constant (1) if one of the black holes is spinless
as was the case in \cite{levin2008:2}, (2) if the binaries
have exactly equal mass \cite{gopakumar2005}, or (3) if both spins are
aligned or anti-aligned with the angular momentum.
Case (1) is worked out thoroughly in paper I \cite{levin2008:2}.
To see that the claim is true in the equal mass case (2), notice that 
$\bs =\delta_1(\bsone+\bstwo)$ and therefore
$\bj=\bl+\bs/\delta_1$. Consequently, $\bj \cdot \bl=L^2+\bs\cdot
\bl/\delta_1$ is conserved since both terms on the right hand side are
conserved. Furthermore, ${\bf \dot{{\bf S}}_{\rm eff}}=\delta_1({\bf \dot{{\bf S}}_1} +{\bf \dot{{\bf S}}_2})$
and it follows from Eq.\ (\ref{dL}) that the change in $\bs$ is always
perpendicular to $\bs$ so its magnitude remains constant. In case
(3), when the spins are aligned or anti-aligned with the orbital
angular momentum, motion is confined to a plane and there is no
precession. Therefore $\bs$ and $\bl$ are constants and the rest follows.

The results of this paper will apply to a general
$\bs$ unless explicitly stated otherwise.

\begin{figure}
\hspace{-30pt}
  \centering
\includegraphics[width=85mm]{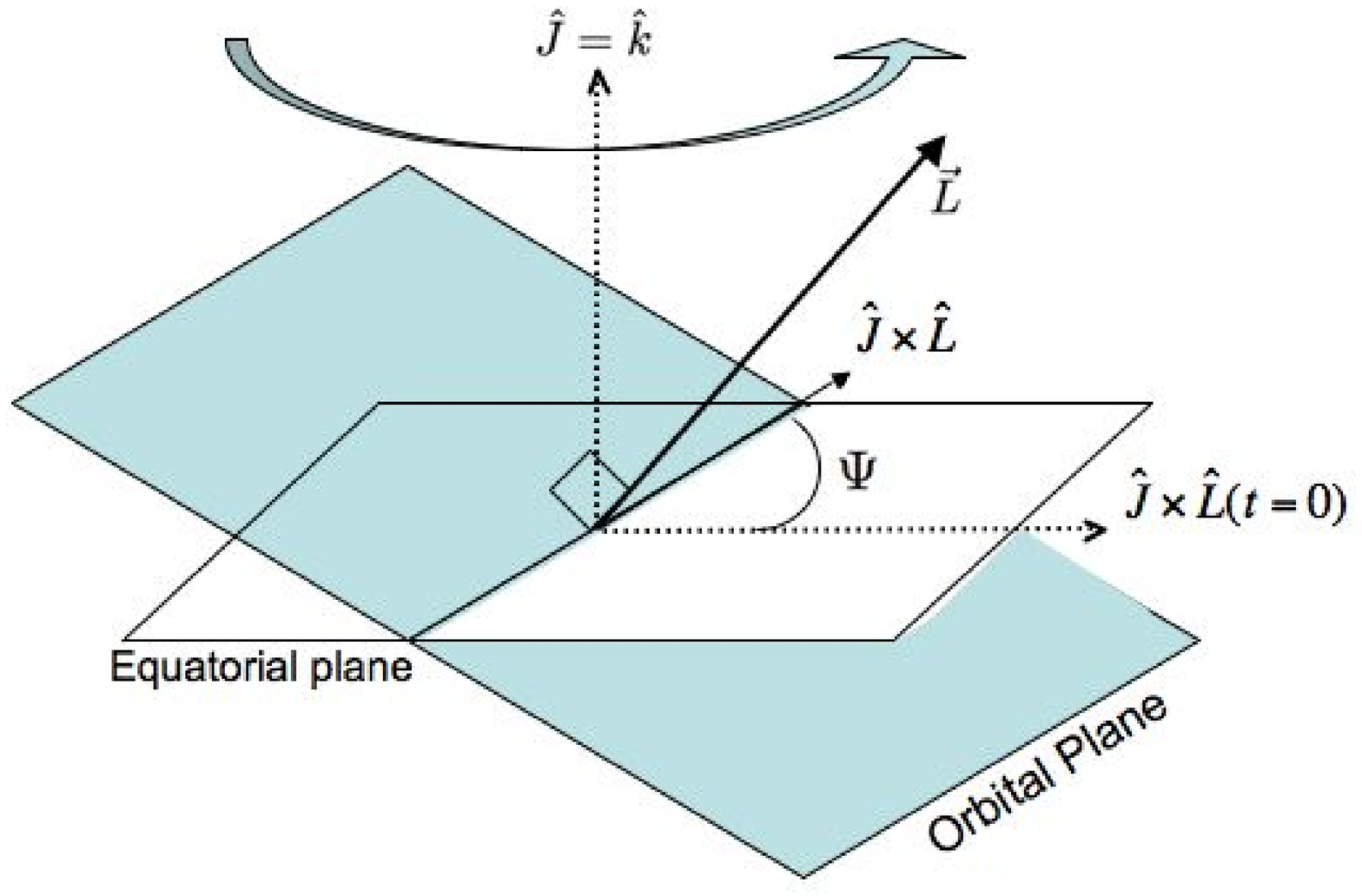}
\includegraphics[width=80mm]{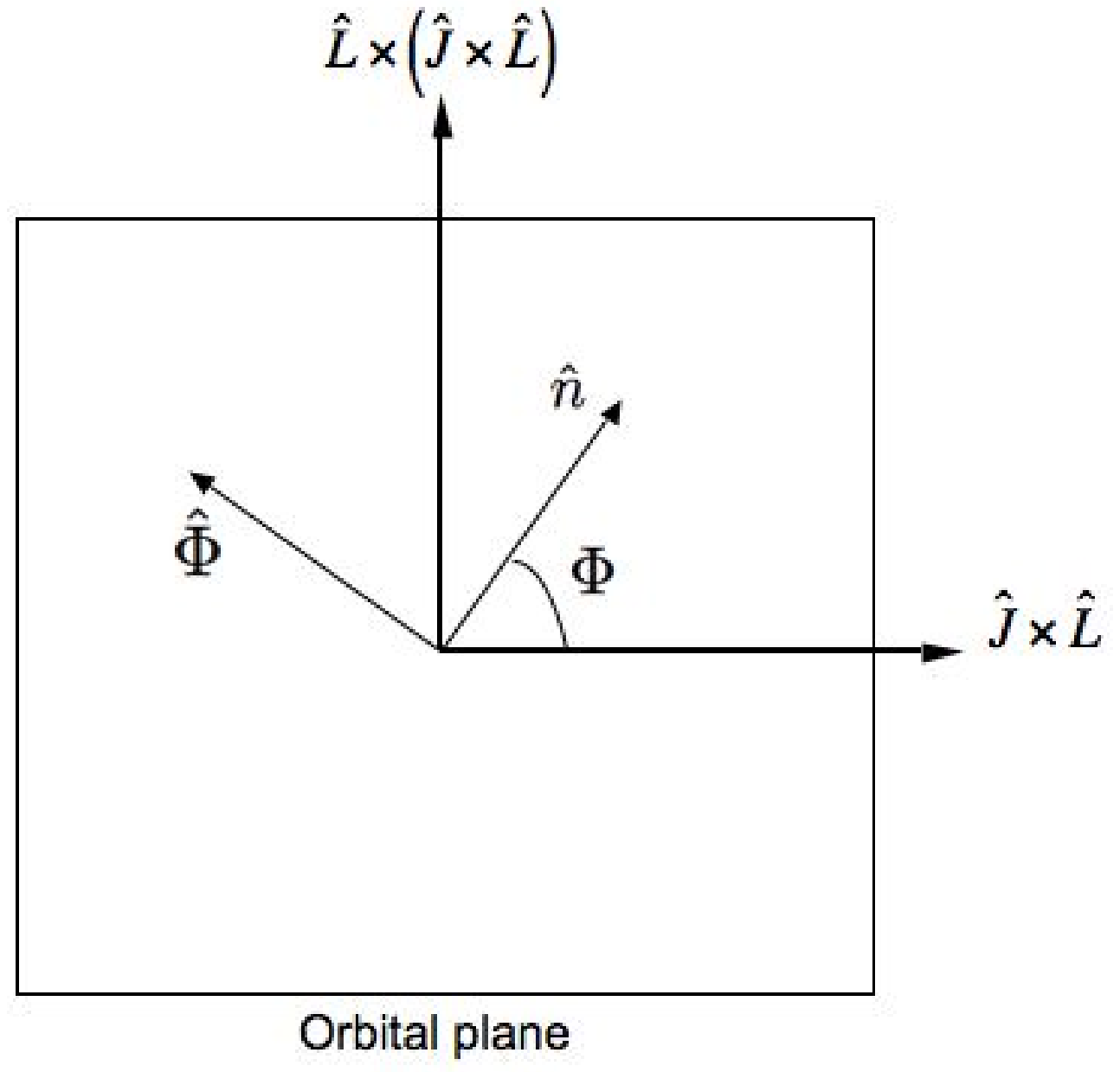}
\hfill
  \caption{Upper: The orbital plane precesses around the $\bhj=\bhk$
  axis through the angle $\Psi$.
Lower: The orbital plane can be spanned by the vectors
$(\bn,\bhPhi)$.
  \label{orbplane2}}
\end{figure}

\subsection{Equations of Motion in the Orbital Plane}
\label{eomsec}

In a non-orthogonal orbital basis, the equations of motion assume a
simple form that allows us to analyze the dynamics of the black hole
pairs. 
The plane perpendicular to the precessing orbital angular momentum 
is spanned by the vectors $(\bn,\bhPhi)$ where
$\bn=\br/r$ and 
\begin{equation}
\bhPhi=\bhl\times\bn \quad .
\end{equation}
Notice $\bn\cdot \bhPhi=0$ so these basis vectors are orthogonal. The
entire orbital plane then precesses around the constant total angular
momentum $\bj=\bl+\bsone+\bstwo$ in the direction $\bhPsi$ defined through 
\begin{equation}
\bhPsi =\bhj\times\frac{(\bhj\times \bhl)}{\left |\bhj\times
  \bhl\right |} \quad .
\end{equation}
The construction, familiar from classical celestial mechanics 
\cite{{schafer1993},{wex1998},{gong2004},{konigsdorffer2005}}, is
illustrated in Fig.\ \ref{orbplane2}.
Incidentally, this basis is explicitly constructed for $\bj \times \bl \ne
0$. When the spin and orbital angular momentum are aligned, anti-aligned, or spin
is zero then $\bj\times\bl=0$, motion is confined to a plane, and we should use the usual
equatorial planar basis.

Notice that $\bhPsi$ is not orthogonal to the orbital plane and
therefore our orbital basis $(\bn,\bhPhi,\bhPsi)$ is not orthogonal.

From Eq.\ (\ref{hameoms}), we can find equations of motion in
coordinates $(r,\Phi,\Psi)$ and their canonical momenta
$(P_r,P_\Phi,P_\Psi)$. As in appendix \ref{orbitalapp}, where we follow
the approach of paper I, it is
convenient to first isolate
the equations of motion in the orbital plane
for the variables $(r,\Phi)$ and their canonical momenta $(P_r,P_\Phi)$:
\begin{widetext}
\begin{empheq}[box=\fbox]{alignat=2}
\label{eoms}
\dot r& =AP_r +B\,,\qquad& 
\dot P_r& =A\frac{P_\Phi^2}{r^3} +CP_r +D+3P_\Phi\frac{\bs\cdot \bhl}{r^4}\\ \nonumber
\dot \Phi  &=A \frac{P_\Phi}{r^2}
+\frac{\bs\cdot \bhl}{r^3}-\dot\Psi (\bhj\cdot\bhl)
\,,\qquad&
\dot {P}_\Phi& =0
\end{empheq}
\end{widetext}
where $A,B,C,D$ are functions of $(r,P_r)$ to be defined momentarily.
The momentum $P_\Phi=L$ conjugate to $\Phi$ is conserved, 
by the justification in paper I which survives despite the
addition of a second spin.
In any other basis, although $L$
is constant, it is not a momentum conjugate to any coordinate. Instead, $L$ should
be interpreted in terms of the linearly independent coordinates and
momenta appropriate for that basis. The added beauty of
this non-orthogonal approach is that $L$ {\it is} a canonical momentum, namely
$P_\Phi$, not so in the usual spherical coordinate basis where $L$ is
neither a coordinate nor a momentum and the Hamiltonian 
angular equations of motion are less transparent.

The four Eqs.\ (\ref{eoms}) describe motion within the orbital
plane. The orbital plane itself precesses around the constant $\bhj$
with variable rate $\dot \Psi$, derived in appendix \ref{dotpsiapp} to be
\begin{equation}
\dot \Psi=\left (\frac{\bhj \times \left (\bs\times \bhl\right)}{\left
  |\bhj\times \bhl\right |r^3}\right )\cdot \bhPsi \, , \quad 
\dot P_\Psi=\left (\frac{\bs\times \bhl}{r3}\right )\cdot \bhj \quad ,
\label{dotPsi}
\end{equation}
and $P_\Psi=\bl\cdot \bhj=L_z$ is not a constant.

Again, using the manipulations of 
appendix \ref{orbitalapp}, particularly Eq.\ (\ref{ap:use}) 
and the identities $\bhj\cdot \bhl=\cos\theta_L=P_\Psi/P_\Phi$,
$|\bhj\times\bhl|=\sin\theta_L$,
we can write the final term in the $\dot \Phi$ equation of
(\ref{eoms}) as 
\begin{equation}
\dot\Psi (\bhj\cdot \bhl)=\left (\frac{\left (\bs\cdot\bhj\right
  )-\left (\bs\cdot \bhl\right )P_\Psi/P_\Phi }{
\left (1-(P_\Psi/P_\Phi)^2\right )r^3}
\right )\frac{P_\Psi}{P_\Phi} \quad .
\end{equation}
Writing it in this form exploits the dependences on the 
coordinates, conjugate momenta, and the constant
$\bs\cdot\bhl$. The one term that clearly remains dependent on angles
is the term
$\bs\cdot\bj$.
Therefore, when both black holes spin, the angular equations will depend on
the angular precession of the orbital plane. 

We saw in paper I a dramatic 
simplification in the case of one effective spin. 
As follows from the earlier discussion of the constants of motion,
$\bs\cdot\bhj$ would be constant if either of the spins vanished and
this 
would remove the angular
depedence in the above equations. 
A pair of spinning black holes of equal mass is also reducible to a
system with effectively one spin (see
 \S \ref{oneff}). We continue to consider the general case of two
 misaligned spins for arbitrary mass ratios.

For completeness, and as a complement to paper I, we
explicitly write out the functions 
$A,B,C,D$ that 
were set up in \cite{levin2008:2} as derivatives on the
Hamiltonian: 
\begin{widetext}
\begin{align}
A =1+&
\frac{1}{2}\left (3\eta -1 \right ) \bp^{2}-\left (  3+\eta
\right)\frac{1}{r}+
\nonumber \\
&
\frac{3}{8} \left(1-5\eta+5{\eta}^{2} \right)
\left(\bp^{2}\right)^{2} 
 +\frac{1}{4}\left[ 2\left(5-20\eta-3{\eta}^{2}
   \right)\bp^{2}
-2{\eta}^{2}\left(\bn\cdot\bp\right)^{2}\right]\frac{1}{r} 
 +\left(5+8\eta \right)\frac{1}{r^{2}}
\nonumber \\
&
\frac{1}{16}\left(-5+35\eta - 70{\eta}^2+35{\eta}^{3}
\right)\left(\bp^{2}\right)^{3}+\nonumber \\
& \frac{1}{8}\left[3
\left(-7+42\eta-53\eta^2-5\eta^3 \right)(\bp^{2} )^2
 +
 2\left(2-3\eta \right) {\eta}^{2}\left(\bn\cdot\bp
 \right)^{2}
\bp^{2}+3\left(1-\eta\right){\eta}^2
\left(\bn\cdot\bp \right)^{4}\right ]\frac{1}{r}
+\nonumber \\
&  \left[\frac{1}{4} \left(-27+136\eta+109{\eta}^2 \right)
  \bp^{2} + \frac{1}{8}\left(17+30\eta\right)\eta
  \left(\bn\cdot\bp \right)^2 \right ]\frac{1}{r^2}
+\nonumber \\
&   2 \left[-\frac{25}{8} +\left(\frac{1}{64}{\pi}^{2}
  - \frac{335}{48} \right)\eta -\frac{23}{8}\eta^{2}\right]
\frac{1}{r^3}
\end{align}
\begin{align}
B =
&- \eta \left (\bn\cdot \bp  \right)\frac{1}{r}
+ \nonumber \\
&  
\frac{1}{8}\left[
  -4{\eta}^{2}\left (\bn\cdot\bp\right)\bp^{2}-12{\eta}^{2}\left(\bn\cdot\bp \right)^{3} \right]\frac{1}{r} 
+\nonumber\\
 &3\eta\left(\bn\cdot \bp\right)\frac{1}{r^{2}}
+\nonumber \\
& \frac{1}{16}\left [2\left(2-3\eta \right)
  {\eta}^{2}\left(\bn\cdot\bp
  \right)\left(\bp^{2}
  \right)^{2}+12\left(1-\eta\right){\eta}^2
  \left(\bn\cdot\bp
  \right)^{3}\bp^{2}-30{\eta}^{3}\left(\bn\cdot\bp 
\right)^{5}\right]\frac{1}{r}
+\nonumber
\\
 & \left[\frac{1}{8}\left(17+30\eta\right)\eta \left(\bn\cdot\bp \right)\bp^2 
+ \frac{1}{3}\left(5+43\eta \right)\eta
\left(\bn\cdot\bp \right)^{3}
\right]\frac{1}{r^2}+\nonumber \\
& 2\left(-\frac{85}{16}-\frac{3}{64}{\pi}^{2}
-\frac{7}{4}\eta\right)\eta\left(\bn\cdot\bp
\right)\frac{1}{r^3} \\
C=&-\frac{B}{r}\\
D=&-(\bn\cdot\bp)C
-\frac{1}{r^2}-\frac{1}{2}\left ( \left ( 3+\eta \right)
\bp^{2}+\eta \left (\bn\cdot \bp \right )^{2}
\right)\frac{1}{r^2}+\frac{1}{r^{3}}
+\nonumber \\
&
\frac{1}{8}\left[ \left(5-20\eta-3{\eta}^{2}
  \right)\left(\bp^{2}
  \right)^{2}-2{\eta}^{2}\left(\bn\cdot\bp\right)^{2}\bp^{2}
-3{\eta}^{2}\left(\bn\cdot\bp \right)^{4} \right]\frac{1}{r^2}
+\nonumber \\
 &\left[\left(5+8\eta \right)\left({\bp}^{2} \right)
  +3\eta\left(\bn\cdot
  \bp\right)^{2}\right]\frac{1}{r^3}
-\frac{3}{4}\left(1+3\eta \right)\frac{1}{r^{4}}
+\nonumber \\
&\frac{1}{16}\left[\left(-7+42\eta-53\eta^2-5\eta^3  \right)(\bp)^3
 +\left(2-3\eta \right) {\eta}^{2}\left(\bn\cdot\bp
 \right)^{2}\left(\bp^{2}
 \right)^{2}+3\left(1-\eta\right){\eta}^2
 \left(\bn\cdot\bp
 \right)^{4}\bp^{2}-\right.
\left. 5{\eta}^{3}\left(\bn\cdot\bp
 \right) ^{6}\right]\frac{1}{r^2}
+\nonumber
\\
& 2\left[\frac{1}{16} \left(-27+136\eta+109{\eta}^2 \right)\left(
  \bp^{2}\right)^2 + \frac{1}{16}\left(17+30\eta\right)\eta
  \left(\bn\cdot\bp \right)^2 \bp^2 \right.
+\nonumber
\\
& \left. \frac{1}{12}\left(5+43\eta \right)\eta
\left(\bn\cdot\bp \right)^{4} \right]\frac{1}{r^3}
+3\left \{ \left[-\frac{25}{8} +\left(\frac{1}{64}{\pi}^{2} -
  \frac{335}{48} \right)\eta
  -\frac{23}{8}\eta^{2}\right]\bp^{2} 
+\right.
\nonumber
\\
& \left.\left(-\frac{85}{16}-\frac{3}{64}{\pi}^{2}
-\frac{7}{4}\eta\right)\eta\left(\bn\cdot\bp \right)^{2}
\right\}\frac{1}{r^4}
+4\left[\frac{1}{8}+\left(\frac{109}{12}-\frac{21}{32}{\pi}^{2}
  \right)\eta \right]\frac{1}{r^5}
\label{ABCD}
\end{align}
\end{widetext}
where $\bn\cdot\bp=P_r$ and $\bp^2=P_r^2+L^2/r^2$.
Notice that $A,B,C,D$, which come from the non-spinning part of the
Hamiltonian \cite{levin2008:2}, depend only on $(r,P_r)$ and
constants.

Useful results can be drawn from a simple observation.
The radial equations in (\ref{eoms}) have no angular dependence.
The energy, angular momentum,
and radius of spherical orbits can be derived from the radial equations alone.
Therefore we can find spherical orbits simply despite the precession
of the orbital plane.

The fact that the two equations in $(r,P_r)$ form a self-contained system
is a restatement of
the fact that the Hamiltonian itself can be viewed in a
one-dimensional effective approach as a function of $(r,P_r)$ and constants.
It is important to be cautious however
when investigating the angular motion. The Hamiltonian depends only on 
$(r,P_r)$ and constants in time, yet those constants in time have to
be carefully varied as functions of the angular coordinates and their
conjugate momenta in a given basis to correctly derive the remaining
equations of motion. This accounts for the labor in appendix \ref{orbitalapp}
needed to derive the $(\Phi,P_\Phi)$ and $(\Psi,P_\Psi)$ equations of motion.

Still, the simple dependences of the Hamiltonian 
allows us to analyze the spherical orbits as one-dimensional radial
motion in a simple effective
potential. The location of the spherical orbits was implicit in paper I to
frame the distribution of all other orbits. For completeness we
determine the range of spherical orbits with an eye on that companion work.

\section{Spherical Orbits}
\label{spherical}
\subsection{Effective Potential for Spinning Black Holes}

\begin{figure}
\includegraphics[width=65mm]{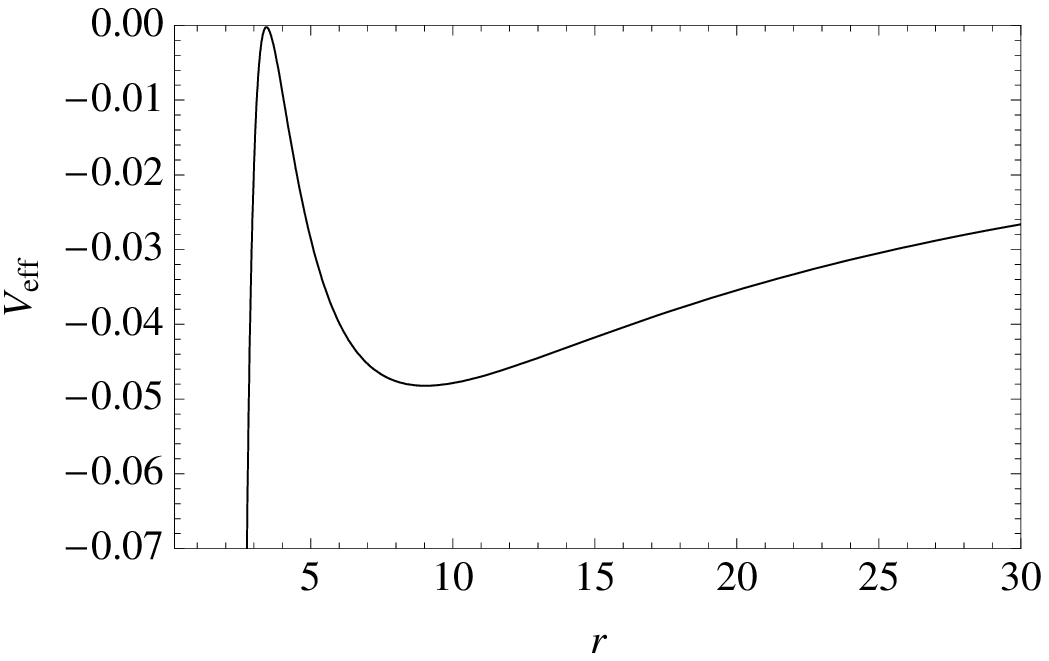}
\includegraphics[width=65mm]{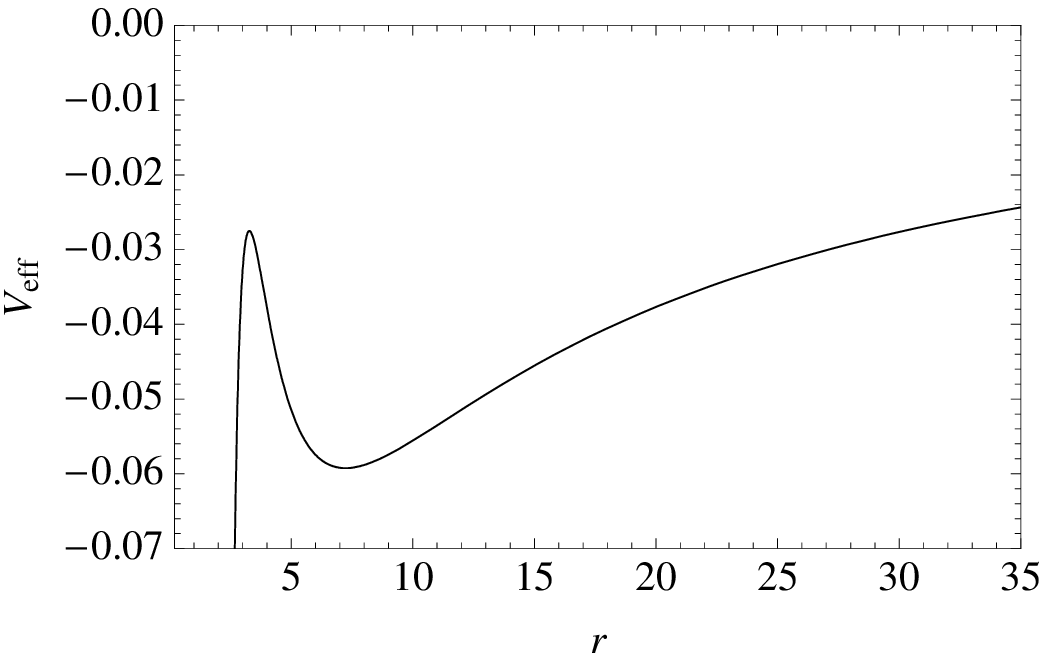}
\includegraphics[width=65mm]{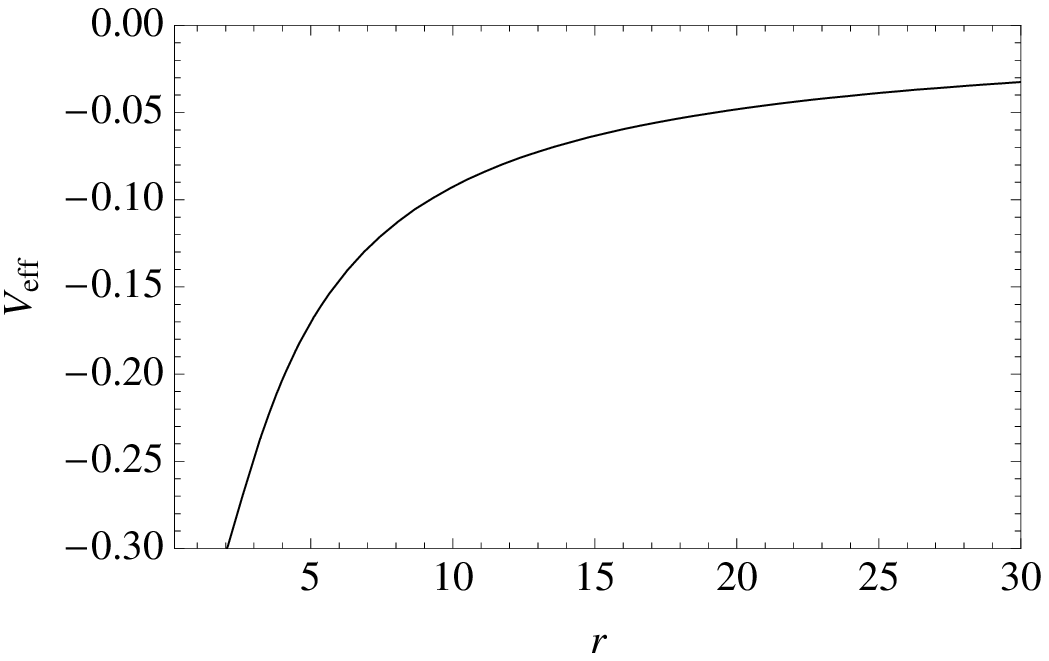}
\includegraphics[width=65mm]{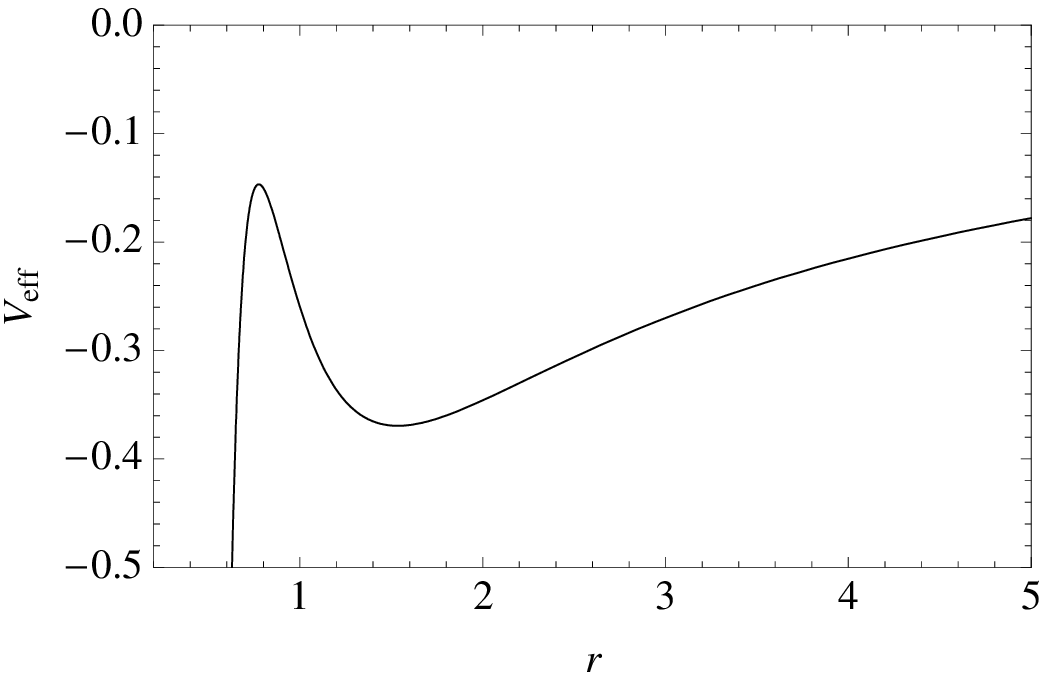}
\hfill
  \caption{An effective potential for two spinning
black holes $a_1=a_2=3/4$ of mass ratio $m_2/m_1=1/4$ for different
values of the angular momentum. 
Notice the
change in scale between panels. 
Upper: The appearance of the ibso
is
marked by the effective potential touching the line $H=0$.
Next: As the angular momentum decreases, the potential will have both
stable and unstable spherical orbits.
Next: As the angular momentum is further decreased there occurs a
critical value at which the unstable and stable spherical orbits merge
at a saddle point, the isso. 
Lower: The last panel shows a difference from the Schwarzschild or Kerr stories. At angular
momenta and radii below the occurence of
the isso, there occur new sets of stable and unstable 
spherical orbits. These occur at radii far below which
the approximation can be trusted, yet we point out their presence for completeness.}
\label{eff}
\end{figure}

Ideally, in an effective potential formulation, the radial equation
could be cast in the form:
\begin{equation}
\half \dot r^2+{\rm effective \ potential}={\rm constant}
\end{equation}
where the effective potential depends only on $r$ and
constants of the motion.
Now, the
Hamiltonian of Eqs.\ (\ref{ham1})-(\ref{Hterms}) does not admit a simple
effective potential formulation since it is a complicated function of
$\bp^2$. We have already argued that $H(r,\bp,\bs)$ can be written as
an effective function of 
$(r,P_r)$ and constants, yet it remains a polynomial function
of $P_r$. However, if we only consider 
\begin{equation}
V_{\rm eff}=H({P_r=0}) \quad ,
\end{equation} 
then we have a good
representation of a pseudo effective-potential {\it at the turning
 points}. We cannot misuse the $V_{\rm eff}$ by trying to interpret
motion away from the turning points, but it gives a perfectly valid
description of the behavior at aphelia and periastra as well as on
spherical orbits. Hereafter we'll shorthand the term ``pseudo
effective-potential'' by ``effective potential''. 

From the Hamiltonian of Eqs.\ (\ref{Hterms}), the effective
potential,
\begin{equation}
V_{\rm eff}(r,L,\bs\cdot\bhl,\eta) 
\quad \quad ,
\end{equation}
is a function of orbital parameters $(r,L,\bs\cdot\bhl)$ and the mass
ratio. Again, since $L$ and $\bs\cdot\bhl$ are
constants of the motion, for a given $(L,\bs\cdot\bhl)$ and a
given mass ratio, the potential is
a function of $r$ only.

Fig.\ \ref{eff} shows several snapshots 
taken of the effective potential for
a pair of spinning black holes as the magnitude of $L$
decreases for a given $\bs\cdot \bhl$ value. 
(For a detailed exposition on interpreting effective potentials for
black hole orbits see Refs.\ \cite{wald1984} and \cite{levin2008:3,perez-giz2008}.)
The spherical orbits are simply the extrema of the 
potential.\footnote{Orbits with the same angular momentum as a stable spherical orbit but
different energy will oscillate
between two turning points, both of which can
be read off the effective potential diagram. Again, due to spin
precession, for misaligned spins these eccentric orbits lift out of a plane. Their spectra
was shown in paper I for a spin/spinless black hole pair.}
An example of such an orbit was shown in Fig.\
\ref{examplesphere}. Although this orbit is not generally periodic, it
does close in the orbital plane as shown in the lower panel of Fig.\ \ref{examplesphere}.

The top panel of Fig.\ \ref{eff} marks the
value of $L$ for which a margnially bound, unstable spherical orbit
appears. An orbit is marginaly bound if its energy $H=0$ and it is
spherical and unstable if it is a maximum of the effective
potential. The conditions are summarized as
\begin{align}
V_{\rm eff}(P_r=0) &=0 \nonumber \\
\frac{\partial V_{\rm eff}}{\partial r} & =0 \nonumber \\
\frac{\partial^2 V_{\rm eff}}{\partial r^2} & <0 
\quad \quad\quad ({\rm ibso}),
\label{ibsocond}
\end{align}
although the first two are sufficient.
We call the margnially bound unstable spherical orbit 
``ibso'' in analogy with the innermost unstable circular orbit (ibco) of
equatorial orbits. 

For angular momenta below $L_{\text{ibso}}$ there will be
both a stable and unstable, energetically bound spherical orbit, as in
the second snapshot of Fig.\ \ref{eff}, until
the angular momentum gets so low that we reach the third snapshot from
the top. Here, the unstable and stable spherical orbits have merged in
a saddle point, coined an innermost stable spherical orbit (isso):
\begin{align}
\frac{\partial V_{\rm eff}}{\partial r} & =0 \nonumber \\
\frac{\partial^2 V_{\rm eff}}{\partial r^2} & =0 
\quad \quad \quad ({\rm isso}).
\end{align}

The 
story plotted out by panels 1-3 of Fig.\ \ref{eff} for
$L_{\text{isso}}<L<L_{\text{ibso}}$ qualitatively follows the fully relativistic
Schwarzaschild and Kerr stories as expected
\cite{levin2008:3,perez-giz2008}. However, something peculiar then happens in the
PN approximation at very
low values of the angular momentum. New stable and unstable spherical
orbits can appear as shown in the bottom panel of Fig.\ \ref{eff}. Or,
for some other ranges of parameters, the ibso disappears or the isso disappears
or both disappear. Sometimes these problems occur, as in the figure, for radii
far below the confidence of the
PN approximation.
We point out these troublesome features in the spirit
of full disclosure. More than this, the details provide a quantitative
testing ground for the
approximation. 

Despite these oddities at low $r$ where the PN-approximation would
make no claims of quantitative validity anyway, the qualitative features of spherical orbits,
homoclinic orbits, and zoom-whirl behavior should survive improved
approximations and full numerical treatments  \cite{Pretorius2006}.
We will locate the $E$ and $L$ of spherical orbits in the next
subsection.

\subsection{Orbital Parameters for Spherical Orbits}
\label{orbparam}

For a given black hole pair, that is, a given mass ratio and $\bs\cdot\bhl$,
all orbits are uniquely specified by their $(E,L)$. 
Using the effective potential, we can easily generate the $E$ and $L$
for spherical orbits and thereby generate initial data for them. 
Initial conditions for spherical orbits were also found in
\cite{{damoureob2001},{buonanno2006}}. Damour \cite{damoureob2001} noticed
that when only spin-orbit terms are included that the Hamiltonian
could be expressed as a radial function. One could arrive at this
conclusion, as we have in the previous section.\footnote{
However, in another basis such as the usual basis for spherical coordinates
 used in \cite{buonanno2006}, 
projection of the vector equations
  of motion will give equations of motion in
 component from that
continue to 
  depend on angles even when only one black hole spins, unlike the
  orbital basis of \S \ref{oneff}.}  The
constant radius orbits occur at the extrema of $H(r,P_r)$,
  in the same spirit as an effective potential method.

\begin{figure}
  \centering
\includegraphics[width=75mm]{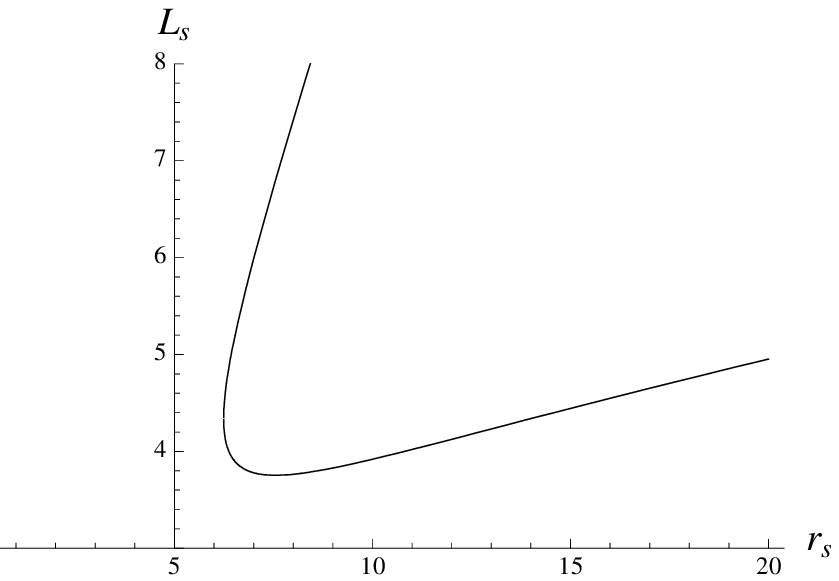}
\hspace{+15pt}
\includegraphics[width=75mm]{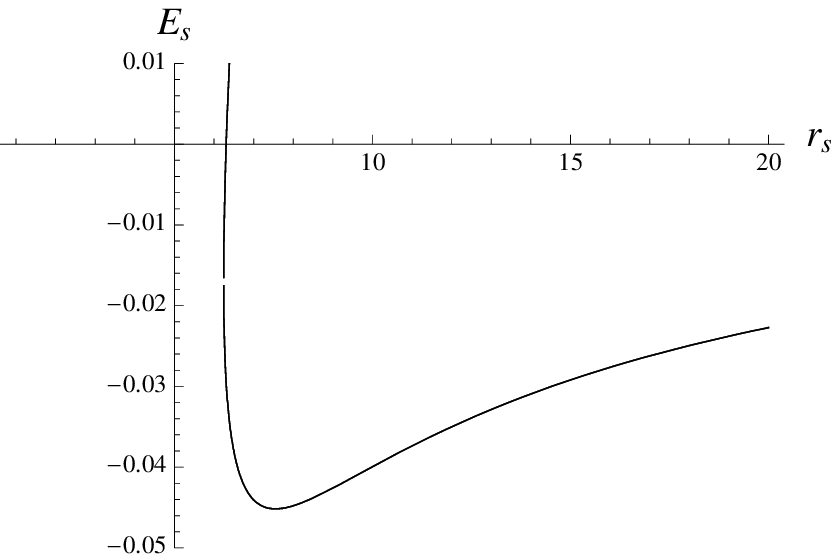}
\hfill
  \caption{$(m_2/m_1=10^{-6},S_{\rm eff}=0)$. Upper:
Angular momentum vs $r_s$.
Lower: Energy vs $r_s$.}
\label{Lc}
\end{figure}

From the vantage point of the effective potential,
spherical orbits satisfy
the condition
\begin{equation}
\frac{\partial V_{\rm eff}}{\partial r} =0
\label{veffzero}
\end{equation}
treating $L$ and $\bs\cdot\bhl$ as constants. We could also
  take the vantage point of the equations of motion.
Since, $P_r=0$ forces $B=0$, the condition
$\dot r=AP_r+B=0$ can be thought of as synonymous with the condition that $P_r=0$.
The constant radius condition is thus the requirement that
\begin{equation}
\left.\dot P_r\right |_{P_r=0}=0
\label{prdotzero}
\end{equation} 
in Eqs.\ (\ref{eoms}) and Eq.\ (\ref{prdotzero}) is equivalent to
Eq.\ (\ref{veffzero}).

The
roots of the 8th-order equation in $L$, Eq.\ (\ref{prdotzero}), give the angular momenta of
the spherical orbits as a function of spherical radius,
$L_s(r_s)$, where we use a subscript $s$ to denote
a quantity evaluated at a spherical orbit. (When there is no spin, the condition reduces to a quartic
in $L^2$ with only two of the four roots real.) Piecing together the real
roots we find $L_s$'s such as the one in Fig.\ \ref{Lc}. 
Although the upper branch grows very quickly in $L_s$, these values rapidly
become physically unreachable since it would require
angular velocities greater than the speed
of light to be that high up on the upper branch. 
One can think of $(L_s/r_s)< 1$ as a crude marker of 
physically allowed values.
To find the energy of spherical orbits, $E_s(r_s)$, we simply plug
$L_s(r_s)$ into the Hamiltonian when $P_r=0$. The energy plot is also
shown in Fig.\ \ref{Lc}.

There are several things to notice about the $L_s$ and $E_s$ plots,
for which we chose values to illustrate the PN 
approximation to Schwarzschild $(m_2/m_1= 10^{-6},S_{\rm
  eff}=0)$.\footnote{Since the radial equation depends only on the
  combination $\bs\cdot\bhl$, Fig.\ \ref{Lc} should be equally valid
  for a non-zero effective spin that is orthogonal $\bl$.}
The large values
of $r_s$ correspond to stable spherical orbits. (Since spin is
zero, these constant radius orbits are actually circular 
equatorial but we'll
keep the language more general.) When $L_s$ hits a
minimum, we have found the isso -- for no $L<L_{\text{isso}}$ are there spherical
orbits.
To the left of that minimum are the unstable spherical orbits. 
A true peculiarity of the figure is the fact that the radii of the unstable spherical
orbits begin to move out to larger $r$.
This is simply a flaw in the PN approximation and does not occur in the
Schwarzschild system. In the fully relativistic system the unstable spherical
orbits always move to smaller radii than the isco, hence the ibco is
really innermost, earning its name. Here, the ibso is not actually innermost -- due to
the poor quality of the approximation -- although it
remains the highest energy bound spherical orbit when it exists. The
ibso cannot be read off of Fig.\ \ref{Lc}
although it can be found simply as the coincident of the roots of
Eqs.\ (\ref{ibsocond}).

Figs.\ \ref{Lc} are for a non-spinning extreme-mass-ratio binary and
are therefore valid as an approximation to Schwarzschild. The
details of these figures will be useful for a future test of the PN
expansion. Here we note that $L_{\text{ibso}}\approx 4.69$, which is about $17$\% higher
than the Schwarzschild value of $4$ while $L_{\text{isso}}\approx 3.75$, which is about
$8$\% higher than the Schwarzschild value of $\sqrt{12}$. The
energy of the ibso is designed to be zero so is not informative but the
energy of the isso is $E_{\text{isso}}\approx -0.0452$, which is about $21$\%
less negative, that is less
energetically bound, than the Schwarzschild case
$(2\sqrt{2}/3)-1$.  Due to the
approximate nature of the expansion it is not necessary to take these
comparisons to heart, but they indicate how the spherical orbits and
the periodic spectra could facillitate a test of the PN expansion.
For a comparison of the isso in different PN approaches including the
resummed Kerr-like effective-one-body approach see \cite{damourpn2000:2,buonanno2006}.

\begin{figure}
  \centering
\includegraphics[width=75mm]{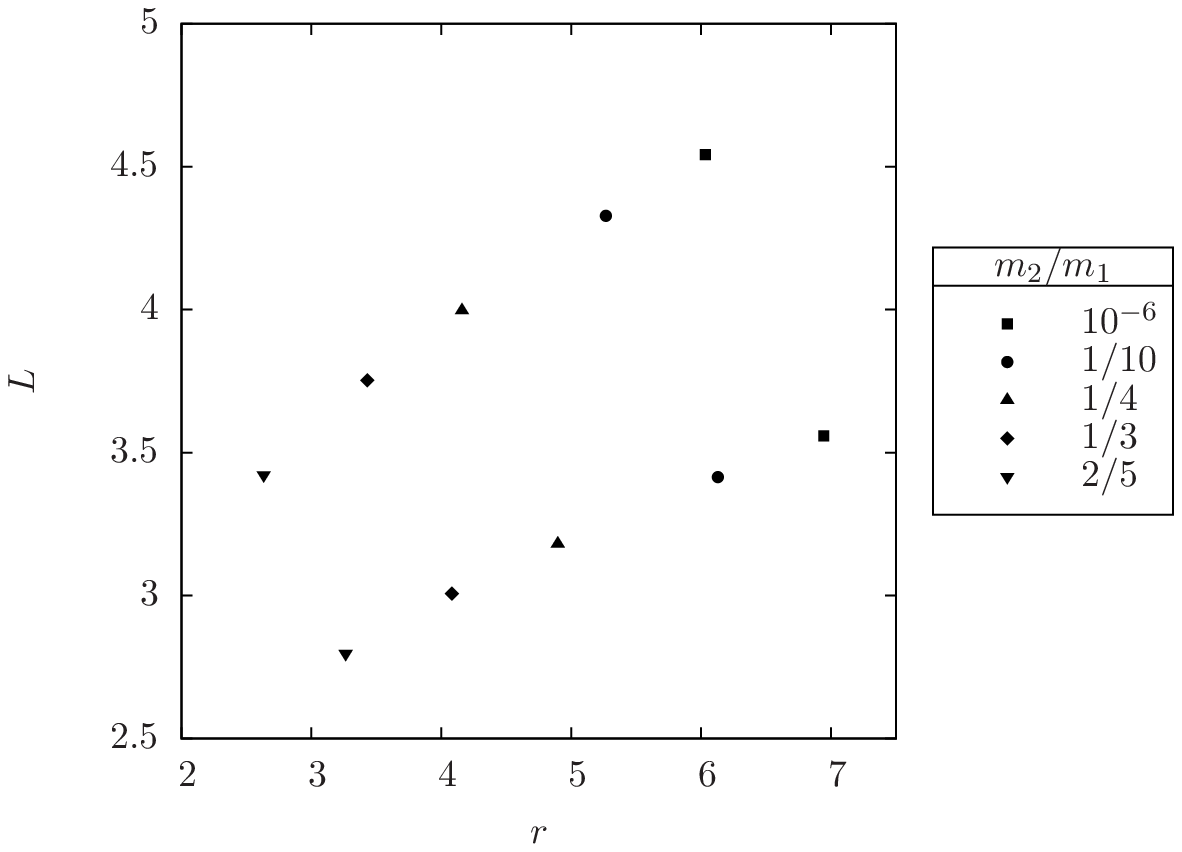}
\includegraphics[width=75mm]{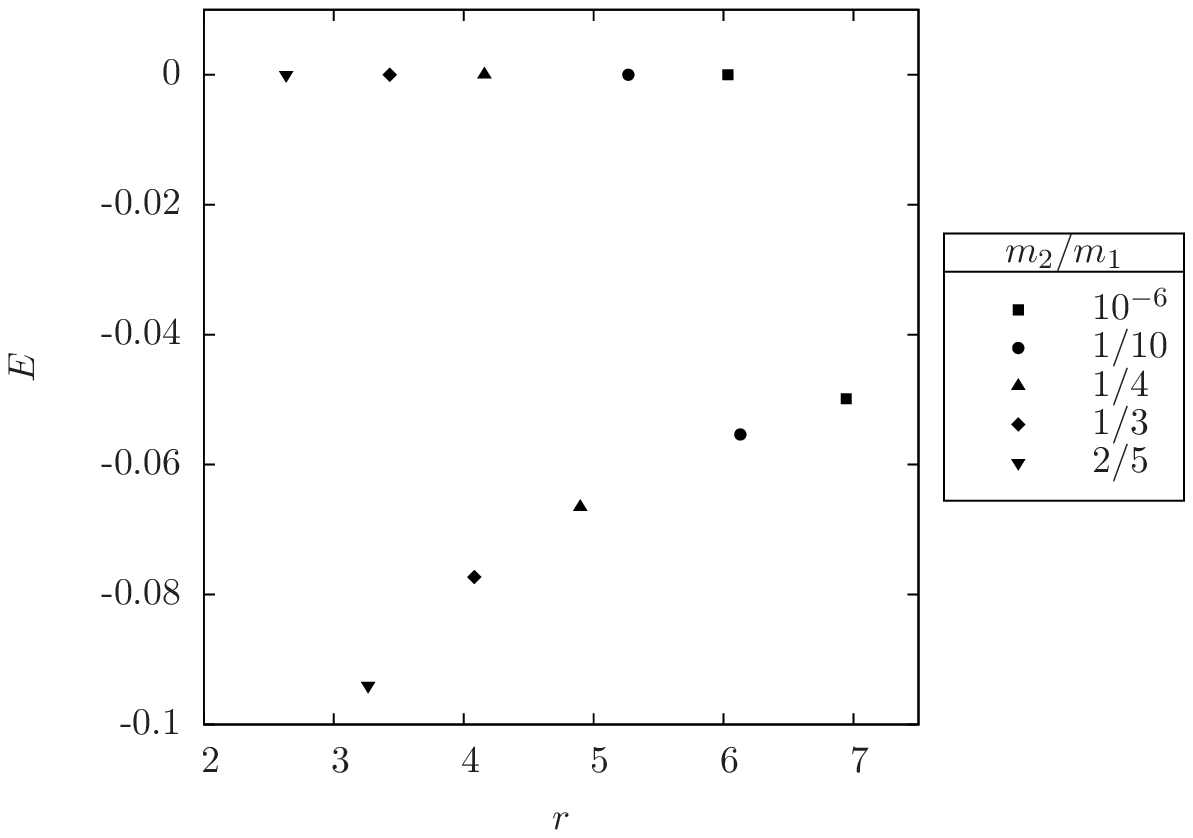}
\hfill
  \caption{All black hole pairs represented have
    $\bs\cdot\bhl=0.35355$.
Upper: Angular momentum vs $r$ for the ibso and isso for
  different mass ratios. The upper point is always the ibso
  for a given symbol while the lower point with the same symbol is
  always the isso. The key lists the different $(m_2/m_1)$. Lower: Energy vs $r$. }
\label{pro}
\end{figure}

\begin{figure}
  \centering
\includegraphics[width=75mm]{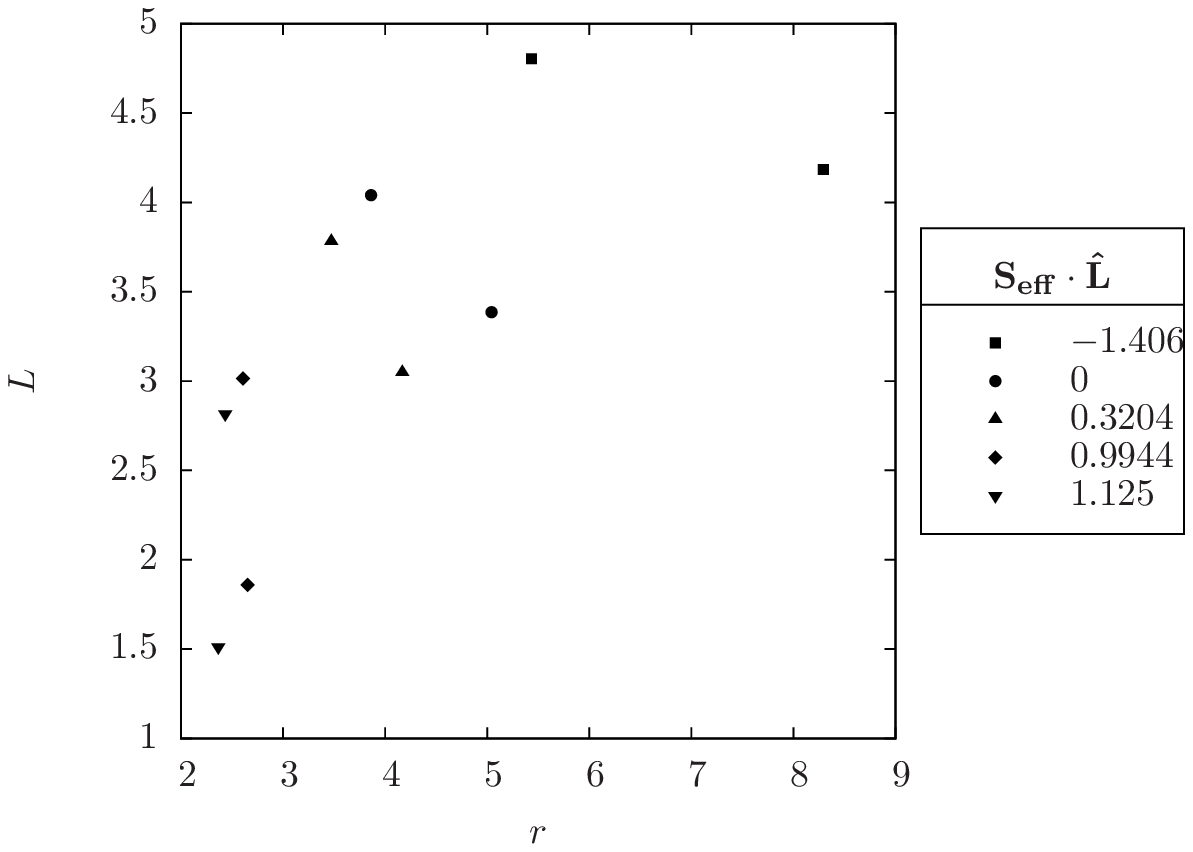}
\hspace{+20pt}
\includegraphics[width=75mm]{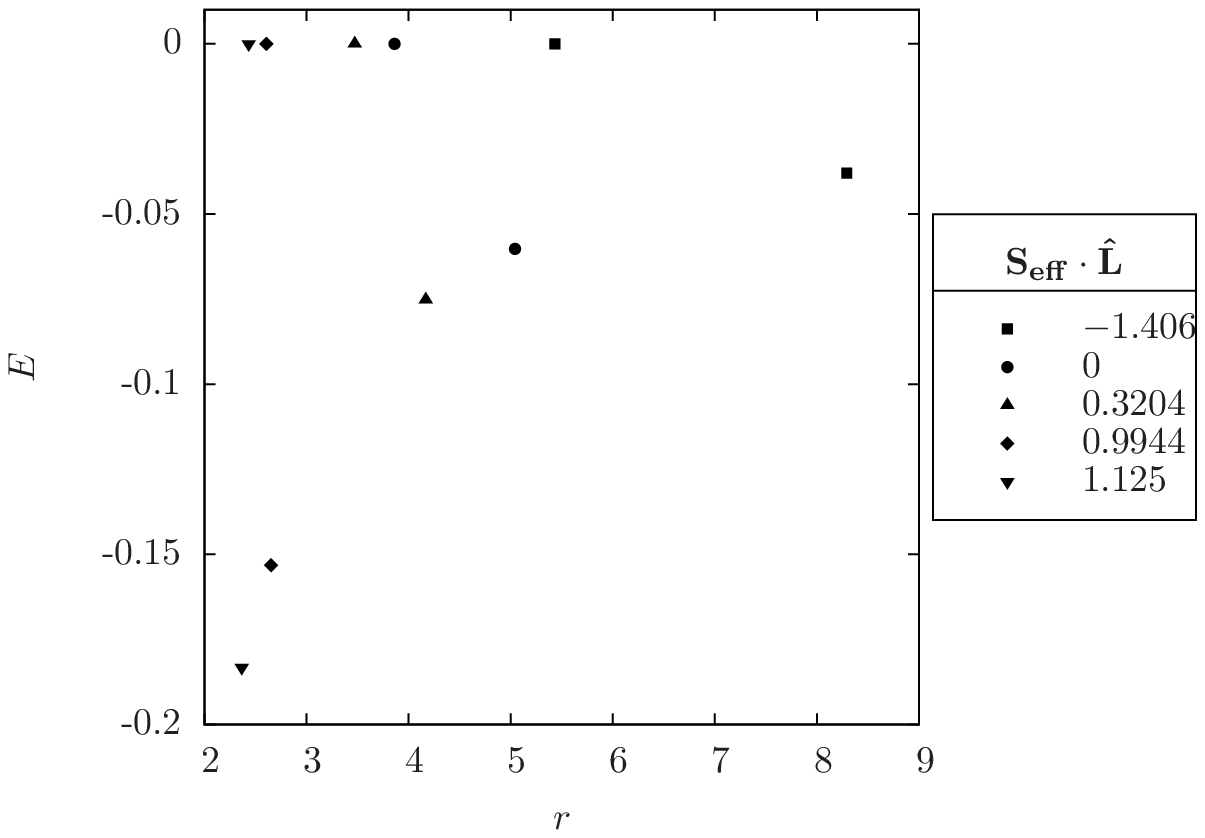}
\hfill
  \caption{Upper: Angular momentum vs $r$ for the ibso and isso for
  fixed mass ratio $m_1/m_1=1/3$ but varying $\bs\cdot\bhl$. 
The upper point is always the ibso
  for a given symbol while the lower point with the same symbol is
  always the isso. The key lists $\bs\cdot\bhl$. 
Lower: Energy vs $r$.}
\label{retro}
\end{figure}

\subsection{Dependence of Binding Energies on mass ratios and spin}

For completeness, we can see how the ibso and the isso vary as the
mass ratio and spins of the
black holes are varied; that is, as their mass ratio and spins are varied. 
For one, since the ibso and isso frame
the distribution of orbits, they define the ranges of $E$ and $L$
values for all other orbits in the strong-field. 
For another, the energy at the isso gives an estimate of 
the energy emitted on quasi-circular inspiral
up to the transition to plunge.
A larger binding energy at the isso
    could also mean a larger signal at final coalescence so these
    variations attest to various levels of detectability. We will
    discuss the binding energy of the isso in this section. In the
    next section we will consider the transition to plunge for
    eccentric orbits.

A black hole pair is specified by its mass ratio, $m_2/m_1$, and its
spins through the
particular combination $\bs\cdot\bhl$. The Hamiltonian, and the radial
equations, depend only on these two combinations.
We will therefore consider the variations
in the isso and ibso for black hole pairs distinguished only by their
 $(m_2/m_1,\bs\cdot\bhl)$ values.
It is important to realize that there is a great deal of degeneracy
among pairs. The ibso and isso values (their energy,
angular momenta, and radial values) are identical for two physically
distinct black hole pairs. 
For instance, a black hole with mass ratio $m_2/m_1=1/3$ and $\bs\cdot
\bhl=0.35355$ could be a black hole with initial values
$a_1=1/4$,
$a_2=0$,
and the spin of the heavier black hole aligned with the initial orbital
angular momentum. However, this is not the only combination of spin
amplitudes and angles that will give the combination $\bs\cdot
\bhl=0.35355$. While different black hole pairs can give degenerate
isso and ibso values, they will be physically distinguishable through
their angular motion.

In Fig.\ \ref{pro} the $L$ 
of the ibso and isso is plotted in the upper panel and the 
$E$ of the isso
and of the ibso are plotted in the lower panel.
Qualitative conclusions can be drawn from these figures. We notice 
    that as the mass ratio is increased towards 1, the radius of both the isso
    and ibso decrease, although the isso moves in faster. Therefore the
    isso is pushed to larger binding
    energies as the mass ratio is increased towards 1.
Because the Hamiltonian is a high-order polynomial in $r$, there can
be more than one marginally bound orbit and more than one saddle point 
for a given $(m_2/m_1,\bs\cdot\bhl)$ pair, as demonstrated in the lowest
panel of Fig.\ \ref{eff}. The second occurence of a marginally bound
orbit and/or saddle point appears in the
vicinity of $r\sim 1$ where the approximation is
uninterpretable. (There may even be third occurences.) It is unclear if
there is any physical content to these other stable and unstable
spherical orbits. Fig.\ \ref{pro} plots only the ibso/isso pair for
$r$ values $>2$.

For the value of $\bs\cdot\bhl\sim 0.35355$ used in the figure, either
the ibso or the isso disappears (or both disappear) as $m_2$ approaches
$m_1$. There
may still be very small radii $(r\sim 1)$ ibso's and/or isso's, but
the sensible ones disappear.
This peculiarity is probably an
artefact of the approximation, a point we return to momentarily. 

Fig.\ \ref{retro} fixes the mass ratio at $m_2/m_1=1/3$ and varies $\bs\cdot\bhl$.
Increasing $\bs\cdot\bhl$ has the same effect of pushing the
isso to smaller separations and therefore to larger
binding energies, although again the isso moves in
faster -- discounting any
marginally bound spherical orbit or saddle points that occur in the
vicinity of $r\sim1$. 
(In fact, as
Fig.\ \ref{retro} shows, at some point the isso actually occurs at a smaller radius
than the ibso.) So, all other factors being equal, spins
anti-aligned with the orbital angular momentum push the isso
out to larger radii and smaller binding energies while aligned spins
pull the isso into smaller radii and larger binding energies.
For the mass ratio of this figure, the ibso or the isso actually vanishes (or
both vanish) as $\bs\cdot\bhl$ is increased much beyond the values shown.

These trends are
consistent with those for spherical orbits discussed in
Ref.\  \cite{buonanno2006}. For 
the equal mass case with spins aligned or
anti-aligned with the orbital angular momentum, the authors of that reference also
remarked on the absense of an isso (called a last stable spherical
orbit (lsso) in their lexicon). Indeed, they used this failing to
argue that the PN expansion could not be used to study the transition
from inspirl to plunge and advocated instead the seemingly more
reliable effective-one-body (EOB) approach
\cite{buonanno1999,damoureob2001,buonanno2007}.
It would be interesting to extend to the EOB method the investigation of the zoom-whirl
orbits of paper I \cite{levin2008:2} and the homoclinic limit of the
zoom-whirls that we turn to in the next section. We leave that to a
future work and continue to use the 3PN Hamiltonian as an example of
our general
method. 

The disappearance of the ibso accompanies the disappearance of all
unstable spherical orbits. Once this happens, there can be no isso
since the isso is really the point of merger of the unstable and
stable spherical orbits.\footnote{This is not the only reason the isso
  disappears. Sometimes the isso disappears because the potential
  simply never flattens out.}
As is already known, in the absence of spin
there are no bound unstable circular orbits at
2PN \cite{kidder1993,wex1993,schafer1993,Damour:1997ub,buonanno1999}. At 3PN there are no bound unstable circular
orbits for mass ratios bigger than about
$m_2/m_1\sim 1/2$. (See also \cite{{damourpn2000:2},{blanchet2002}}.)
The absence of a bound unstable circular orbit is clearly a
shortcoming of the approximation since we know that the Schwarzschild
spacetime possess an unstable circular orbit as a reflection of its
high non-linearity.
Furthermore, the unstable circular orbits
are present in fully relativistic
treatments as the equal mass numerical investigation of
Ref.\ \cite{Pretorius2006}
shows. Therefore, the ibso and isso should emerge for $m_2\rightarrow m_1$ at higher orders.
Incidentally,
their disappearance at 3PN-order implies the expansion is very likely
approximating the dynamics as more stable than it really is and
therefore less vulnerable to chaos than it really is for these
comparable mass binaries.

We have already warned caution to take the  trends as qualitative indicators and not
to invest too much in the numbers due to pressures on the PN expansion
at such large values of $(m_1+m_2)/r$. Afterall, the PN expansion is
an expansion in small $(m_1+m_2)/r$ and will naturally begin to faulter
for small $r$.

We have focused on the binding energy of the isso primarily to fit
into the wider conversation that has focused on quasi-circular
inspiral. However, the eccentric binaries formed by tidal capture in
dense regions will not transition from inspiral to plunge through the
isso. Rather they will transition through the eccentric separatrix
between bound and plunging orbits. We investigate that separatrix
briefly in the final section.

\section{Homoclinic Orbits -- the separatrix between bound and
  plunging orbits}
\label{homoclinic}

It is worthwhile to mention another important kind of orbit that
occurs in our dynamical system, the homoclinic orbit \cite{{bombelli1992},{levino2000}}. 
Homoclinic orbits are intriguing for several reasons, not least of which is that they mark
the orbits through which the transition from inspiral to plunge should occur. In fact, the isso
itself, the transition point for
quasi-circular orbits,
is a zero eccentricity homoclinic orbit
\cite{bombelli1992,levin2008:3,perez-giz2008}. 
We make the connection between the energetically bound, unstable
spherical orbits and the homoclinic orbit explicit in this final section.

Formally,
homoclinic orbits are defined as trajectories that asymptote
to the same hyperbolic invariant set in the infinite future as in
the inifinite past. In these black hole settings, the role of the
hyperbolic invariant set is played by the energetically bound,
unstable spherical orbits. 
Although in the lexicon of black hole physics these orbits have been
coined ``unstable'', they are strictly speaking hyperoblic, which is
to say they possess both a stable eigendirection and an unstable
eigendirection under linear perturbations. And, the eigendirections
lie along the homoclinic orbit in the local neighborhood of the
unstable circle they approach. Although we won't demonstrate that line
up here, the point was emphasized in detail in
Refs.\ \cite{levin2008:3,perez-giz2008} for Kerr equatorial dynamics.

The stability exponents can be found by linearizing in small perturbations  around Eqs.\
(\ref{eoms}). This was done for equatorial Kerr orbits in
Ref.\ \cite{perez-giz2008}. Although we will not write out the
explicit procedure here, we mention that the
spherical orbits have radial eigenvalues that come in
plus/minus pairs, as they must in a Hamiltonian system.
The radial eigenvalues are real for the unstable
spherical orbits and imaginary for the stable spherical orbits. The isso occurs at
the merger of the eigenvalues at zero.
A direct computation of the stability exponents
around circular orbits confirms that the stable spherical
orbits and the unstable spherical orbits are distributed around
the isso as
Fig.\ \ref{Lc} shows.

\begin{figure}
\hspace{0pt}
  \centering
\includegraphics[width=70mm]{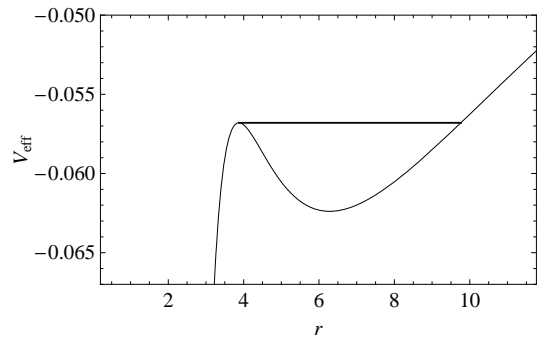}
\hfill
  \caption{An effective-potential for $m_2/m_1=1/4$, with the spin
    of the heavier black hole displaced from $\bhl $ by $\pi/4$
and amplitude    $a_1=1/2$ while the lighter black hole has no spin.
The straight line is the energy of the unstable spherical
orbit. It is also the energy at another, larger turning point $r_a\sim
10$, which identifies the apaastron of the homoclinc orbit.
}
  \label{effone}
\end{figure}

Through the phase space analysis we have shown that the energetically
bound, unstable
circular orbits are actually hyperbolic -- they have a positive
stability exponent as well as a negative stability exponent. We could
compute the eigenvectors and show they lie along the homoclinic orbit
in the local neighborhood of the unstable circle as we did for Kerr in
Ref.\ \cite{perez-giz2008}. However, for our purposes it is sufficient
and illuminating to consider a physical space picture.

\begin{figure*}
\hspace{0pt}
  \centering
\includegraphics[width=60mm]{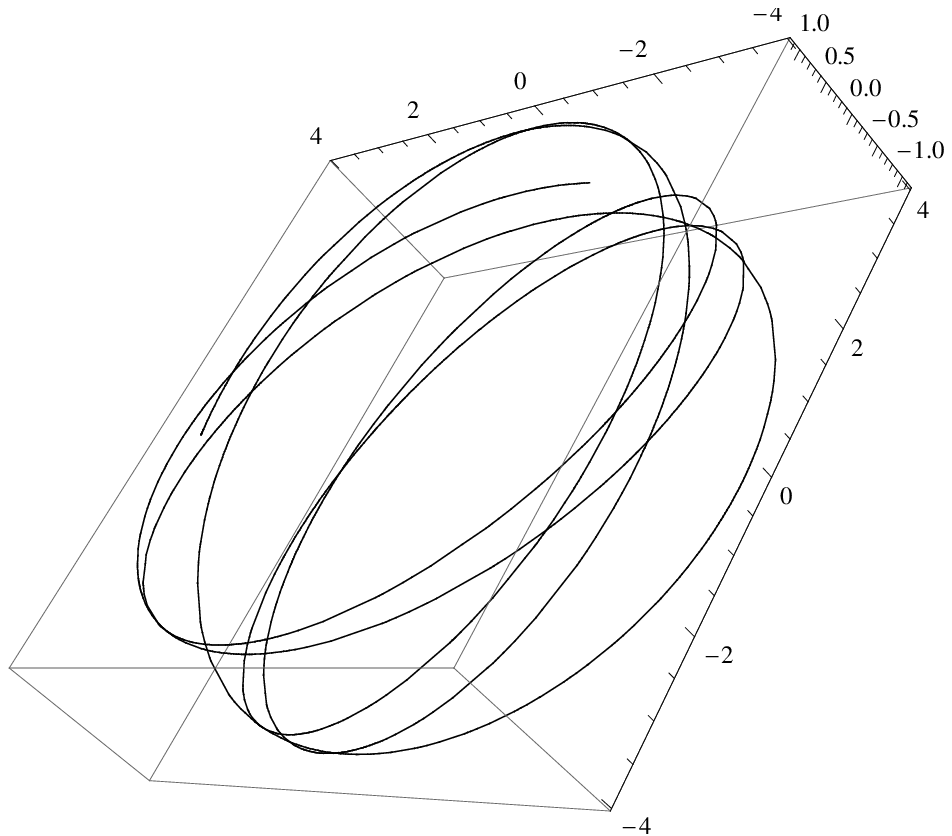}
\includegraphics[width=45mm]{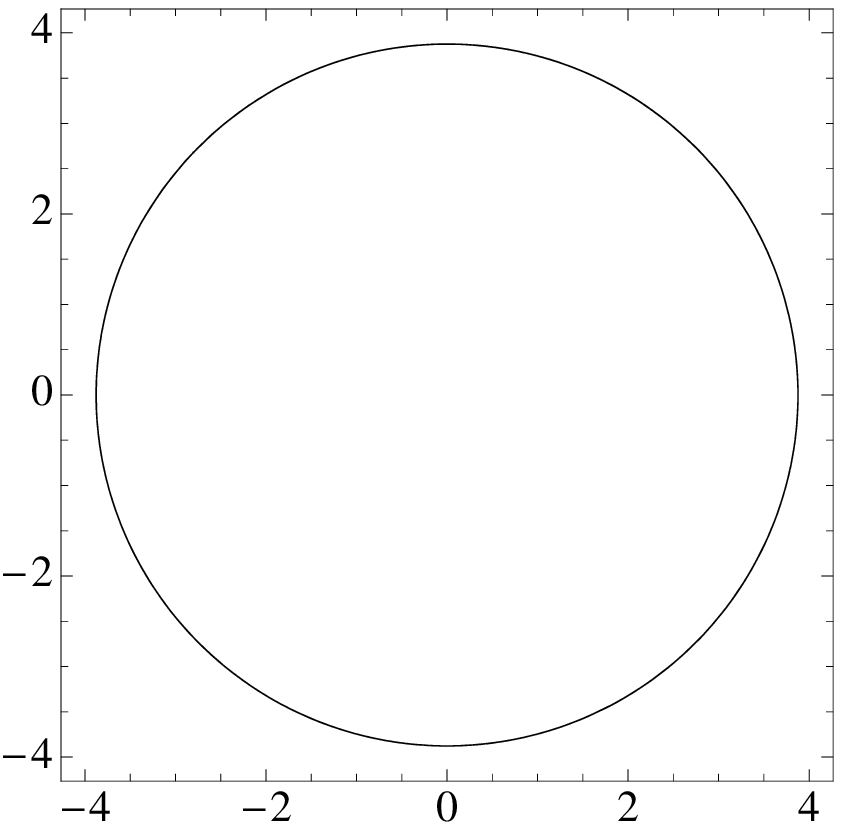}
\hfill
  \caption{The unstable spherical orbit that is the maximum of
    $V_{\rm eff}$ for Fig.\ \ref{effone}. Unlike the effective
    potential, the details of the full orbit do depend on the specific
    combination $\bs\cdot \bhl$. Left: As viewed in three dimensions. Right: As
    viewed in the orbital plane.
}
  \label{unstablephys}
\end{figure*}

We can identify the separatrix -- i.e. the homoclinic orbit -- in an 
effective-potential picture.
In
particular, consider
a binary with mass ratio $m_2/m_1=1/4$ and the heavier black hole
spins with 
amplitude $a_1=1/2$ offset from $\bhl$ by $\pi/4$ while the lighter
black hole is nonspinning. Although, again, any equivalent combination
of
$\bs\cdot\bhl$ is described by this same figure.
The unstable spherical orbit, $r_{u}$, at the maximum of $V_{\rm
  eff}$ in Fig.\ \ref{effone} is drawn in physical space in Fig.\
\ref{unstablephys}. 
Although the orbit is a closed circle in the
orbital plane, it fills out a band on a sphere in three
dimensions. Because of numerical instability near this orbit, we only
show a few windings.

\begin{figure*}
\hspace{0pt}
  \centering
\includegraphics[width=60mm]{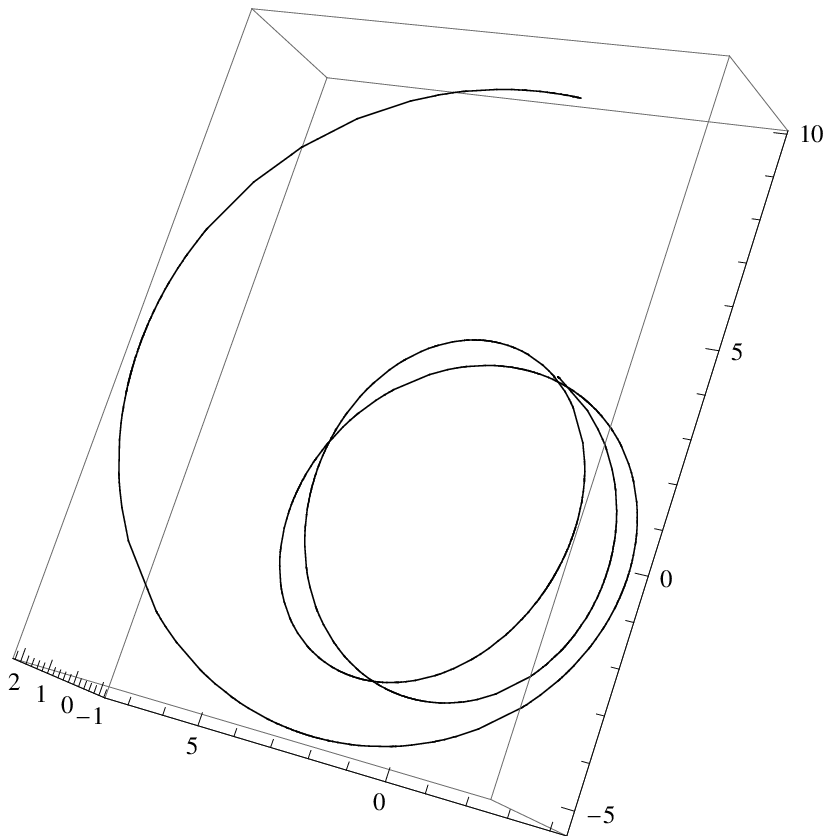}
\includegraphics[width=45mm]{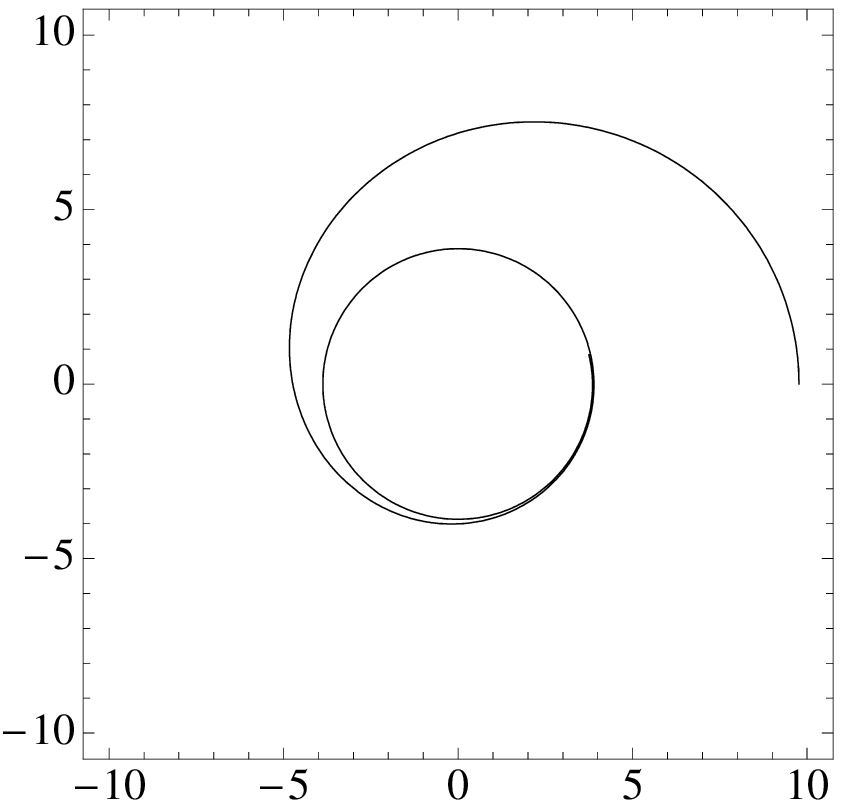}
\hfill
  \caption{The homoclinic orbit for Fig.\ \ref{effone}
approaching the unstable spherical
  orbit. Left: As viewed in 3d. Right: As viewed in the orbital plane.
Because of numerical instability near the highly unstable constant
radius orbit, we only show a few windings.
}
  \label{horb}
\end{figure*}

The energy of this orbit, $E_s(r_u)$, is indicated by a straight
line across the potential of Fig.\ \ref{effone}. Note that this energy
touches the potential at the unstable radius $r_u$ and at
some larger radius, roughly $r\sim 10$.
This larger radius is the apastron of an orbit. If
the two black holes are released from rest at an initial separation
in center-of-mass coordinates
equal to this apastron, their orbit will roll down the
potential (although the shape changes when $P_r\ne 0$) and
then climb back up the other side asymptotically approaching the
spherical orbit at the top of the hill. By definition, this is a
homoclinic orbit (Fig.\ \ref{horb}). To our knowledge it is the first
of its kind to be
found out of the equatorial plane \cite{{levin2008},levin2008:3,{perez-giz2008}}. 

The orbit winds around
the center of mass an infinite number of times as it asymptotically
approaches the unstable spherical orbit. Although not strictly
periodic -- the homoclinic orbit never returns to apastron -- it will be significant for
the periodic tables \cite{levin2008,levin2008:2} as a maximum energy orbit for a given $L$ in
the strong-field regime \cite{levin2008:3,perez-giz2008}.

\section{Conclusions}
\label{sum}

This paper, as the second in a series, provides the energetic frame in
which the periodic tables of paper I were set
\cite{levin2008:2}. Although in support of paper I's goals, the
analysis of spherical orbits could be
relevant to additional tests of the PN expansion for spinning black
hole pairs and could have a place in the disucssion of initial
values for numerical relativity. Additionally, we find the
non-equatorial homoclinic orbits that whirl an infinite number of
times as they asymptote to the unstable spherical orbit. The
homoclinic spearatrix is important as defining the transition to
plunge for all orbits, including eccentric and precessing orbits. It
would be interesting to extend this study to the EOB method 
\cite{buonanno1999,damoureob2001,buonanno2007} in a
future work.

In closing, we comment on an intriguing implication of the set of
spherical orbits for spinning black hole pairs. In this work, we
restricted ourselves to spin-orbit coupling and although we allowed
both black holes to spin, we found that the spherical orbits constrain the
range of allowed bound orbits in the following sense.
For a given angular momentum, spin initial conditions, and mass ratio,
the stable spherical orbit is the lowest energy geodesic and the unstable
spherical orbit is the highest energy orbit in the strong-field --
barring the failures of the approximation at these close
separations.\footnote{We actually consider the emergence of an ibco to
  define the strong-field. For the equal mass cases that resist the
  development of an ibco, it is as if the approximation is not
  effective enough to enter the strong field.} As we
showed in paper I \cite{levin2008:2}, between
these two spherical orbits lies an infinite set of orbits that are
closed in the orbital plane. The periodic set corresponds to a subset
of the rationals, with the rational identifying a given orbit 
increasing monotonically between the stable spherical orbit and the
unstable spherical orbit. The homoclinic orbit is the infinite whirl
limit of the periodic set and would be the final entry in a periodic table
of orbits corresponding to the infinite limit of the rationals. 

This pattern of a periodic set framed by constant radius orbits and
limiting to the homoclinic is consistent with a picture that has
emerged for Kerr black hole orbits
\cite{levin2008,levin2008:3,perez-giz2008}. The consistency of the
picture for two spining comparable mass black holes with the Kerr case
is precisely what
is surprising, or at least intriguing. The geodesics in a Kerr
spacetime are known to be integrable \cite{carter1968}. There are
enough constants of the motion to restrict trajectories to regular
tori and prohibit chaotic mixing. As Poincar\'e intuited, the
structure of the periodic orbits encodes the entire dynamics and the
regularity of the system is in fact reflected in the regularity of the
periodic spectrum. The simplicity of the spherical orbits and
the periodic set they frame
suggests that even when both black holes spin
and are of comparable mass, there is no chaos -- at least not in
physically plausible regimes --if only spin-orbit
coupling is included.\footnote{It is possible that for $\bs\cdot\bhl$
  much larger than black hole physical values chaos could develop.
Afterall, one of the constants of
  motion
$P_\Psi$ has been lost with the inclusion of spin-orbit coupling
  opening the door for chaos.}

Put another way,
homoclinic orbits are also a sign of
non-linearity. They mark the intersection of the stable and unstable
manifolds of a hyperbolic invariant set. They are the precursor to
chaos in the sense that 
under perturbation, the homoclinic orbit breaks up into a homoclinic
tangle and will be the locus of a fractal set of orbits
\cite{ott,perez-giz2008}. The fractal
set is sometimes refered to as a
strange repellor and is the analog for conservative systems of strange
attractors in dissipative systems \cite{cornish1994,dettman1994,Cornish:1996yg,levin2000}. 

Systems with a regular set of periodic orbits that culminate in a
homoclinic limit are not chaotic. 
However, the spinning pairs are vulnerable to chaos as evidenced by
their very possesion of a homoclinic orbit. Indeed
chaos has by now been well confirmed in the
form of a fractal set when spin-spin coupling is included
\cite{levin2000,levin2003,{gopakumar2005},{levin2006},{wu2007},{wu2008}}.
As suspected in Ref.\ \cite{hartl2005}, our work suggests that the
emergence of chaos must be directly tracable to the spin-spin
coupling. 
We conjecture that the transition to chaos could be witnessed through
the destruction of the corresondence of the periodic set with the
rationals when spin-spin coupling is turned on.
The additional precessional effects of spin-spin coupling, we suggest,
must destroy the homoclinic orbit, replace it with a homoclinic tangle
-- a fractal set of orbits -- and induce chaotic scattering among
geodesics in the vicinity.

\vfill\eject
\acknowledgements

We are especially grateful to Gabe Perez-Giz for his valuable and generous
contributions to this work and to Jamie Rollins for his careful
reading of the manuscript.
We also thank
Szabi Marka for important discussions.
This material is based in part upon work
supported under a National Science Foundation Graduate Research
Fellowship. 

\vfill\eject
\vfill\eject
\appendix
\section{Projection the equations of motion onto the non-orthogonal
  orbital basis}
\label{orbitalapp}

\subsection{The Orbital Plane Equations}

The four equations of motion in the orbital plane are obtained by
projecting Hamilton's equations onto the basis vectors, as is done in
celestial mechanics. For now,
consider only the projections onto the orbital basis vectors to
generate the four equations,
\begin{align}
\dot{\br}\cdot \bn&=\frac{\partial H}{\partial {\bp}} \cdot \bn \nonumber \\
\dot{\br}\cdot \bhPhi&=\frac{\partial H}{\partial {\bp}} \cdot \bhPhi\nonumber \\
\dot{\bp}\cdot \bn&= -\frac{\partial{H}}{\partial {\br}}\cdot \bn \nonumber \\
\dot{\bp}\cdot \bhPhi&= -\frac{\partial{H}}{\partial {\br}}\cdot \bhPhi
\quad .
\label{eomshort2pre}
\end{align}
To break down the LHS involves
\begin{align}
\label{ap:step}
\dot{\br}&=\dot r \bn +r {\bf \dot{\bn}} \nonumber \\
\dot{\bp}&=\dot P_r \bn +P_r {\bf \dot{\bn}}-\frac{L}{r^2}\dot r
\bhPhi+\frac{L}{r}{\bf \dot{\bhPhi}} \quad .
\end{align}
We will need projections of ${\bf \dot{\bn }}$ and ${\bf
  \dot{\bhPhi}}$ along $\bn$ and $\bhPhi$.
Now, since $\bn\cdot\bn=1$, it follows that ${\bf \dot{\bn}}\cdot
\bn=0$ and by the same reasoning ${\bf \dot{\bhPhi}}\cdot \bhPhi=0$.
Also, by orthogonality,
\begin{align}
{\bf\dot{\left ({\bn\cdot\bhPhi}\right )}}&=0 \implies \nonumber \\
{\bf \dot{\bn}}\cdot
\bhPhi &=- {\bf \dot{\bhPhi}}\cdot \bn \quad .
\end{align}
To obtain
the final dot product above we 
expand the basis vectors $(\bn,\bhPhi,\bhPsi)$
in terms of an intermediate basis $(\bhx,\bhy)$ that spans the 
orbital plane, and 
then expanding $(\bhx,\bhy)$ in the Cartesian basis. We proceed as we
did in paper I and define the intersection of the orbital plane with
the equatorial plane:
\begin{equation}
\bhx=\frac{\bhj\times \bhl}{\left |\bhj\times \bhl\right |}
=\frac{\bhj\times \bhl}{\sin\theta_L} \quad ,
\label{xhat}
\end{equation}
where $\cos\theta_L=\bhl\cdot\bhj$.
The vector orthogonal to $\bhx$ that lies in the orbital plane is 
\begin{equation}
\bhy=\bhl\times \bhx\quad .
\label{yhat}
\end{equation}
This intermediate orbital basis will be useful in the manipulations
that follow. 
In terms of Cartesian components defined with $\bhk=\bhj$ and
$\bhi,\bhjy$ spanning the equatorial plane, 
we can expand
\begin{eqnarray}
\bhx &=& \cos\Psi \bhi +\sin\Psi \bhjy\nonumber \\
\bhy &=& \sin\theta_Y(-\sin\Psi \bhi +\cos\Psi \bhjy)+\cos\theta_Y\bhk \quad ,
\label{need1}
\end{eqnarray}
where $\cos\theta_Y=\bhy\cdot \bhj$. Since $\bhy$ is always orthogonal
to $\bhl$, again by construction, this is not really a new angle but
can be recast as $\theta_Y=\pi/2 - \theta_L$.

Our non-orthogonal basis can then be expanded as
\begin{eqnarray}
\bn &=&\cos\Phi \bhx+\sin\Phi \bhy \nonumber \\
\bhPhi &=& -\sin\Phi \bhx +\cos \Phi \bhy \nonumber \\
\bhPsi &=& -\sin\Psi\bhi+\cos\Psi\bhjy \quad .
\label{need}
\end{eqnarray}
Using
\begin{eqnarray}
{\bf \dot{\bhx} }&=& \dot \Psi\bhPsi \\
{\bf \dot{\bhPsi}} &=& -\dot\Psi\bhx  \\
{\bf \dot{\bhy}} &=& -\sin\theta_Y\dot\Psi \bhx 
+\cos\theta_Y\dot\theta_Y\bhPsi-\sin\theta_Y\dot\theta_Y\bhk \, \, .
\label{dots}
\end{eqnarray}
From all of the above relations we obtain for use in the projections
\begin{eqnarray}
{\bf \dot{\bn} }\cdot \bn&=& 0\\
{\bf \dot{\bhPhi}} \cdot \bhPhi &=& 0 \\
{\bf \dot{\bn}} \cdot \bhPhi &=& \dot\Phi+\dot\Psi\sin\theta_Y=
\dot\Phi+\dot\Psi\cos\theta_L \\
{\bf \dot{\bhPhi}} \cdot \bn&=& - {\bf \dot{\bn}} \cdot \bhPhi 
\quad .
\end{eqnarray}
Conveniently, these are the same projections we found in paper I for
the case in which only one black hole spins and $\dot \theta_Y=\dot \theta_L=0$.

Now we can derive the equations of motion in the $(r,\Phi,\Psi)$
coordinates.
We use the equations we constructed in paper I \cite{levin2008:2}
\begin{align}
\dot{\br}&=A\bp +B\bn + \frac{\bs\times \br}{r^3} \nonumber \\
\dot{\bp}&=C\bp +D\bn +  \frac{\bs\times \bp}{r^3} 
+3\frac{\bl\cdot \bs}{r^4} \bn
\quad ,
\label{eomshort2}
\end{align}
where $A,B,C,D$ are given by Eqs.\ (\ref{ABCD}).
With the projections (Eqs.\ (\ref{eomshort2pre})), (\ref{ap:step}), and 
the above vector relations we have the radial equation from
$\dot{\br}\cdot \bn$
in (\ref{eomshort2}):
\begin{equation}
\dot r= AP_r +B \quad .
\end{equation}
The $\Phi$ equation follows from
\begin{eqnarray}
\dot{\br}\cdot \bhPhi &=&\frac{\partial H}{\partial \bp}\cdot \bhPhi \\
r\left (\dot \Phi +\dot\Psi\cos\theta_L\right ) &=&
A \frac{L}{r}+
 \frac{\left (\bs\times \br\right
  )\cdot \bhPhi}{r^3} \quad .
\label{rgamma}
\end{eqnarray}
Look at
\begin{eqnarray}
\left (\bs\times \br\right  )\cdot \bhPhi &=& 
r\left (\bs\cdot \bhl \right )\quad .
\end{eqnarray}
The $\Phi$ equation is then
\begin{eqnarray}
\dot \Phi
 &=& A\frac{L}{r^2}
-\dot\Psi \cos\theta_L+\frac{\bs\cdot \bhl}{r^3}
\label{group}
\end{eqnarray}
where $\bhs\cdot \bhl$ is constant.

The two conjugate momenta equations are next.
We start with $P_r$:
\begin{eqnarray}
\dot {\bp}\cdot \bn 
&=&\dot P_r
-\frac{L}{r}(\dot\Phi+\dot\Psi\cos\theta_L) \\
&= & 
 C P_r +D 
+2
\frac{\bs\cdot \bl}{r^4} 
\nonumber
\end{eqnarray}
where we have used that
\begin{eqnarray}
(\bp \times \bs)\cdot \bn &=& \frac{\bs \cdot \bl}{r} 
\end{eqnarray}
Notice if we use Eq.\ (\ref{group}), we have 
\begin{eqnarray}
\dot P_r =
A\frac{L^2}{r^3}
+  
 C P_r +D +3\frac{\bs\cdot \bl}{r^4} 
\nonumber
\end{eqnarray}
and last
\begin{eqnarray}
\dot {\bp} \cdot \bhPhi &=& P_r(\dot\Phi+\dot\Psi\cos\theta_L)
-\frac{L}{r^2}\dot r
\\
&=& C \frac{P_\Phi }{r}
+
\frac{P_r \bs\cdot \bhl}{r^3}
\nonumber
\label{pgammaeom}
\end{eqnarray}
where we have used that
\begin{equation}
(\bp \times \bs)\cdot \bhPhi = \bs \cdot (\bhPhi
  \times \bp)=-P_r \bs\cdot \bhl
\end{equation}
Notice if we use Eq.\ (\ref{group}), we have a cancellation and
\begin{eqnarray}
 \left (A P_r -\dot r\right )\frac{L}{r^2} &=& -B\frac{L}{r}
= C \frac{L }{r}
\nonumber
\label{pgammaeom}
\end{eqnarray}
which confirms a true statement but does not provide any new equation of
motion. The final equation of motion is simply 
$\dot P_\Phi =0$.
All four equations in the orbital basis are compiled in the boxed Eqs.\ (\ref{eoms}).

\bigskip

\subsection{The Precession of the Plane}
\label{dotpsiapp}

The plane precesses in the direction $\bhPsi$ at a rate $\dot \Psi$,
which can be computed from the first of Eqs.\ (\ref{dots}):
\begin{equation}
{\bf \dot{\bhx} }= \dot \Psi\bhPsi \quad\quad .\nonumber 
\end{equation}
We can isolate $\dot \Psi$ by projecting along $\bhPsi$,
\begin{equation}
{\bf \dot{\bhx} }\cdot \bhPsi= \dot \Psi \quad\quad .
\end{equation}
We take the time derivative of Eq.\ (\ref{xhat}) and use the constancy
of $\bhj$ and the precession equation for ${\bf \dot{\bhl}}$ from
Eq.\ (\ref{ds2}) to find
\begin{equation}
\dot\Psi=\left (\frac{\bhj\times \left (\bs\times \bhl\right )}{\left
  |\bhj\times \bhl\right |r^3}\right )\cdot \bhPsi \quad .
\end{equation}
Notice that the term that would have been proportional to
$\dot\theta_L$ is killed since it is also proportional to $\bhx\cdot\bhPsi=0$.
With some vector manipulations, including the general rule $\bf
A\times (\bf B\times \bf C)=\bf B (\bf A\cdot \bf C)-\bf C(\bf A\cdot
\bf B)$, applied to both the term in parantheses and to $\bhPsi=\bhj
\times \bhx$ with $\bhx$ given by Eq.\ (\ref{xhat}), this can be reduced to 
\begin{equation}
\dot\Psi=\frac{\bs \cdot \left (\bhj-\bhl(\bhj\cdot\bhl)\right
  )}{\sin\theta_L^2r^3} \quad .
\label{ap:use}
\end{equation}
Going in the other direction with the general rule,
$\bf B (\bf A\cdot \bf C)-\bf C(\bf A\cdot
\bf B)=\bf
A\times (\bf B\times \bf C)$, we can write the right-hand-side as a
triple cross product and identify 
the particularly compact form
\begin{equation}
\dot\Psi=\frac{\bs\cdot \bhy}{\sin\theta_L r^3} \quad .
\label{compact}
\end{equation}
As in paper I, 
\begin{equation}
P_\Psi=L_z =\bl\cdot\bhj\quad \quad .
\end{equation}
Unlike paper I, $P_\Psi$, which can also be expressed as $P_\Psi=L
\cos\theta_L$,
is not conserved when both black holes spin and precess.

\bigskip
\subsection{One Effective Spin}
\label{oneff}

The equations of motion simplify considerably if there is only one
effective spin, such as
the case of only
black hole spinning \cite{levin2008:2}:
\begin{empheq}[box=\fbox]{alignat=2}
\label{eomssimp}\nonumber
\dot r& =AP_r +B\,,\qquad& 
\dot P_r& =A\frac{L^2}{r^3} -\frac{B}{r} P_r +D+3\delta_1\frac{\bsone\cdot \bl}{r^4}\\ \nonumber
\dot {\Phi}& =A\frac{L}{r^2}-\delta_1\frac{L}{r^3}\,,\qquad&
\dot {P}_\Phi& =0\nonumber 
\end{empheq}
The orbital plane
precesses with frequency 
\begin{equation}
\dot\Psi=\Omega_L=\delta_1
\frac{J}{r^3} \quad \quad \dot P_\Psi=0.
\end{equation}
Consequently,
the equations of motion above
are independent of angles. In paper I, we used these purely radial equations to
study several features of the dynamical system, such as a periodic table 
that defined the spectrum of black hole orbits.

The same simplification can be effected when the black holes are of
equal mass $m_1=m_2$. Then what we really mean by $\bsone$ is
$\bsone\rightarrow \bsone+\bstwo$.

\bibliographystyle{aip}
\bibliography{pnspherical}

\end{document}